**Classification: Physical Sciences/Physics**

# Anomalous localization behaviors in disordered pseudospin systems: Beyond the conventional Anderson picture

A. Fang[a, b], Z. Q. Zhang[a, b], Steven G. Louie[b, c, d], and C. T. Chan[a, b, 1]

[a]Department of Physics, The Hong Kong University of Science and Technology, Clear Water Bay, Hong Kong, China; [b]Institute for Advanced Study, The Hong Kong University of Science and Technology, Clear Water Bay, Hong Kong, China; [c]Department of Physics, University of California at Berkeley, Berkeley, California 94720, USA; [d]Materials Sciences Division, Lawrence Berkeley National Laboratory, Berkeley, California 94720, USA

[1]To whom correspondence should be addressed. E-mail: phchan@ust.hk

# Abstract

**We discovered novel Anderson localization behaviors of pseudospin systems in a 1D disordered potential. For a pseudospin-1 system, due to the absence of backscattering under normal incidence and the presence of a conical band structure, the wave localization behaviors are entirely different from those of normal disordered systems. We show both numerically and analytically that there exists a critical strength of random potential ($W_c$), which is equal to the incident energy (*E*), below which the localization length $\xi$ decreases with the random strength *W* for a fixed incident angle $\theta$. But the localization length drops abruptly to a minimum at $W = W_c$ and rises immediately afterwards, which has never been observed in ordinary materials. The incidence angle dependence of the localization length has different asymptotic behaviors in two regions of random strength, with $\xi \propto \sin^{-4}\theta$ when $W < W_c$ and $\xi \propto \sin^{-2}\theta$ when $W > W_c$. Experimentally, for a given disordered sample with a fixed randomness strength *W*, the incident wave with incident energy *E* will experience two different types of localization, depending on whether $E > W$ or $E < W$. The existence of a sharp transition at $E = W$ is due to the emergence of evanescent waves in the systems when $E < W$. Such localization behavior is unique to pseudospin-1 systems. For pseudospin-1/2 systems, there is a minimum localization length as randomness increases, but the transition from decreasing to increasing localization length at the minimum is smooth rather than abrupt. In both decreasing and increasing regions, the $\theta$-dependence of the localization length has the same asymptotic behavior $\xi \propto \sin^{-2}\theta$.**

Anderson localization is one of the most fundamental and universal phenomena in disordered systems. Anderson's seminal work in 1958 [1] has inspired intensive studies on the effect of randomness in a vast variety of electronic and classical wave systems [2-10]. Meanwhile, the rapid progress in experimental techniques enables us to reach an unprecedented level of manipulating artificial materials such as ultracold atomic gases [11] and nano/micro-dielectric structures [12], making it possible to create new materials with unusual transport properties [11-14]. The interplay between disorder and new artificial materials continues to generate many amazing phenomena, such as the suppression of Anderson localization in metamaterials [15-17], supercollimation of electron beams in 1D disorder potentials [18] and delocalization of relativistic Dirac particles in 1D disordered systems [19].

Among these new materials, pseudospin-1/2 materials are of particular interest due to their conical band structure and the chiral nature of the underlying quasiparticle states. A prototypical example of pseudospin-1/2 materials is graphene [13, 14]. The low energy excitations in graphene behave like massless Dirac particles and the orbital wave function can be represented by a two-component spinor, with each component corresponding to the amplitude of the electron wave function on one trigonal sublattice of graphene. We emphasize that the "pseudospin-1/2" in graphene refers to the spatial degree of freedom, and has nothing to do with the intrinsic spin of electrons. In addition to graphene, the Dirac cone and the associated pseudospin-1/2 characteristic of quasiparticles can be found in a wide range of quantum and classical wave systems, such as topological insulators [20-23] and the photonic and phononic counterparts of graphene [24-28]. Recently, pseudospin-1 systems have also attracted a lot of attention [27-42]. Different from the Dirac cones in graphene, a Dirac-like cone is found in pseudospin-1 systems where two cones meet and intersect with an additional flat band at a Dirac-like point [27-42]. For example, certain photonic crystals (PCs) can exhibit such Dirac-like conical dispersions at the center of the Brillouin zone due to the accidental degeneracy of monopole and dipole excitations [27-31]. The physics near the Dirac-like point can be described by an effective spin-orbit Hamiltonian with pseudospin $S=1$ and their wave functions are represented by a three-component spinor. Such systems are called

“pseudospin-1 materials” [31]. These systems have also been theoretically predicted [32-36] and experimentally realized by manipulating ultracold atoms in an optical lattice [37] or arranging an array of optical waveguides in a Lieb lattice [38-41]. As an analogy with gate voltage in graphene and other charged Dirac fermion systems, the potentials in pseudospin-1 systems can be shifted up and down by a simple change of length scale in PCs [31] or an appropriate holographic mask in ultracold systems [32-36]. Due to the specific pseudospin characteristics of the underlying quasiparticles, both pseudospin-1/2 and -1 systems exhibit many unusual transport properties, such as Klein tunneling [14, 31-36] and the related one-way transport in 1D potentials [18, 19, 31]. In 1D disordered graphene superlattices, localization behaviors such as angle dependent electron transmission [43, 44] and directional filtering due to strong angle-dependent localization length [45] have been predicted. We will present some surprising, counterintuitive transport phenomena for pseudospin-1 systems in 1D disordered potentials (see Fig. 1). We will also show results of pseudospin-1/2 systems for comparison.

The Anderson localization behavior in pseudospin-1 systems under 1D disordered potentials is entirely different from that in any disordered systems made of normal materials. In normal disordered materials, it is well known that all states become localized in 1D random potentials due to the constructive interference of two counter-propagating waves in the backward direction [2-5]. However, for pseudospin-1 systems, a disordered 1D potential only gives rise to a random phase in the spatial wave function and does not produce any backward scatterings for waves propagating in the normal direction. Such behavior was first discovered in pseudospin-1/2 systems [14, 19, 46, 47]. In the case of pseudospin-1 electromagnetic waves [31], the absence of backscattering can be interpreted as the impedance match between any two adjacent layers in such systems. Thus, Anderson localization occurs only for obliquely incident waves.

Furthermore, due to the existence of a Dirac-like point, the introduction of a disorder potential makes it possible to have evanescent waves occurring in the system when the potential at certain layer is close to the incident energy. And the presence of evanescent waves also makes the transport of waves entirely different from that in normal disordered systems. In this work, we show both analytically and numerically that, for pseudospin-1 systems, when the randomness is small so that no evanescent waves occur in any layer, the localization length $\xi$ decays with the incident angle $\theta$ according to $\xi \propto \sin^{-4}\theta$ at small $\theta$. However, when the strength of the random potential reaches a critical value, which equals the incident energy of the wave, the localization length drops suddenly to a minimum and rises immediately afterwards as evanescent waves emerge. In the latter case, the $\theta$-dependence of $\xi$ changes to a different behavior, i.e., $\xi \propto \sin^{-2}\theta$. The sudden drop as well as the subsequently immediate rise of $\xi$ with increasing randomness and the change of the asymptotic behavior in the $\theta$-dependence are not seen in any normal disordered systems, to the best of our knowledge (see Fig. 2). In normal disordered systems, $\xi$ always decreases with increasing randomness, consistent with our intuition that disorder should disrupt transmission. The existence of a critical randomness in pseudospin-1 systems suggests some kind sharp transition between two localization phases. The physical origin of such a transition is the occurrence of evanescent waves in certain fluctuating layers with randomness that is beyond the critical randomness. Evanescent waves are known to produce a diffusive-like transport in an ordered graphene at the Dirac point [48, 49]. Here we discover that evanescent waves can produce even more fascinating novel transport behaviors in disordered pseudospin-1 systems. For pseudospin-1/2 systems in 1D disordered potentials, our analytical and numerical results find a smooth crossover in the localization length behavior from a decreasing one at small randomness to an increasing one at large randomness, and an angular dependence of $\xi \propto \sin^{-2}\theta$ in both the localization length decreasing and increasing regimes. As will be shown later, the absence of the sharp transition in pseudospin-1/2 systems is due to the presence of additional interface scatterings, which produces a $\xi \propto \sin^{-2}\theta$ behavior even at small randomness. Thus, the $\theta$-dependent localization length behavior does not change when the randomness is increased.

## Results and Discussions

**Models and Numerical Results.** The systems under investigation are pseudospin-1 systems in 1D disordered potentials, which are in the form of *N* random layers. Each layer has the same thickness $d$, but feels a random potential $V(x)$ with a strength *W*, as shown in Fig. 1. Here, $V=0$ denotes the energy of the Dirac-like point of the background medium. A plane wave is incident on the layered structure at an incident angle $\theta$ from the background with the incident energy $E$. For normal incidence ($\theta=0$), the waves are delocalized, irrespective of the strength of randomness due to the absence of backscattering [31]. Here we consider oblique incidence ($\theta \neq 0$), for which Anderson localization can occur. It has been shown previously that the wave equation of such systems can be described by a generalized 2D Dirac equation with a 1D random potential [31-36],

$$H\psi=[\hbar v_g \vec{\mathrm{S}}\cdot\vec{\mathrm{k}}+V(x)\mathrm{I}]\,\psi=E\psi\,. \tag{1}$$

Here $\psi$ is a spinor function, $\vec{\mathrm{k}}=(k_x,\ k_y)$ is the wavevector operator with $k_x=-i\dfrac{\partial}{\partial x}$ and $k_y=-i\dfrac{\partial}{\partial y}$, $\vec{\mathrm{S}}=(S_x,\ \mathrm{S}_y)$ is the matrix representation of spin-1 operator, $v_g$ is the group velocity, and I is the identity matrix in the pseudospin space. For simplicity, Eq. (1) can be rewritten as

$$[\vec{\mathrm{S}}\cdot\vec{\mathrm{k}}+\bar{V}(x)\mathrm{I}]\,\psi=\bar{E}\psi\,, \tag{2}$$

with $\bar{E}=E/\hbar v_g$ and $\bar{V}(x)=V(x)/\hbar v_g$. The normalized random potential in the *j*-th layer is taken to be $\bar{V}(x)=\mathrm{v}_j$ ($j=1,2,3,\cdots,N$), which is an independent random variable distributed uniformly in the range $[-\bar{W},\ \bar{W}]$ ($\bar{W}=W/\hbar v_g$ is the random strength of the normalized potential). We can calculate the transmission coefficient $T_N$ through a random stack of *N* layers by the transfer-matrix method (TMM) [31]. The localization length $\xi$, or the inverse of the Lyapunov exponent $\gamma$, is obtained through the relation

$$\xi = \gamma^{-1} = -\lim_{N->\infty} \frac{2Nd}{< \ln T_N >_c}, \qquad (3)$$

where $<>_c$ denotes ensemble-averaging.

We first show the localization length as a function of the random strength $\bar{W}$. Results of averaging over 4000 configurations with *N* taken to be five times of the localization length are shown in Figs. 2a and 2b for different incident angles and energies, respectively. At small randomness, these results show that the localization length decreases with increasing randomness following a general form $\xi \propto \bar{W}^{-2}$, similar to the behavior found in ordinary disordered media [3, 4]. However, if $\bar{W}$ is further increased, the localization length $\xi$ drops abruptly to a minimum at a critical $\bar{W}_c = \bar{E}$, independent of incident angle and energy, and rises immediately afterwards.

These results are rather intriguing. First, the cusp-like turnaround of localization behavior is not seen in any other disorder systems to our knowledge. For normal disordered media, $\xi$ always decreases with increasing disorder. Second, the sudden change of localization behavior near the critical random strength $\bar{W}_c = \bar{E}$ indicates some kind sharp transition between two different localization phases: $\bar{W} < \bar{E}$ and $\bar{W} > \bar{E}$ in the $\bar{E}$-$\bar{W}$ space. To further elaborate on this point, we examine the $\theta$-dependence of the localization length. The result of $\bar{E} = 0.02$ and small disorder $\bar{W} = 0.01 (< \bar{E})$ is shown by blue circles in Fig. 3a, where a log-log plot of $\xi$ vs. $\sin\theta$ shows a straight line with a slope of -4 for small incident angles $\theta$, indicating a $\xi \propto \sin^{-4}\theta$ behavior. However, the slope changes to -2 for a higher disorder $\bar{W} = 0.03$ ($> \bar{E}$) (blue diamonds), indicating a $\xi \propto \sin^{-2}\theta$ behavior. There is hence a change of localization behaviors from $\xi \propto \sin^{-4}\theta$ to $\xi \propto \sin^{-2}\theta$ in the two different regions of $\bar{W}$. It will be shown analytically later that this transition occurs exactly at $\bar{W} = \bar{E}$, and the physical origins of the above anomalous localization behaviors are the existence of the Dirac-like point and the occurrence of evanescent waves in some layers caused by a diverging scattering strength when $\bar{W} > \bar{E}$.

To see whether such anomalous localization behaviors also occur in pseudospin-1/2 systems, we studied numerically the localization length behaviors for pseudospin-1/2 systems. The Hamiltonian of pseudospin-1/2 systems has the same form as Eq. (1) except that the wave function is a two-component spinor [14, 18] instead of a 3-component spinor, and the spin matrices become Pauli matrices. The results of the TMM method are shown in Figs. 2c and 2d. Compared to Figs. 2a and 2b, for all incident angles and energies studied, the cusp-like sharp change in $\xi$ does not exist in pseudospin-1/2 systems. Instead, $\xi$ shows a smooth crossover from a decreasing behavior at small randomness to an increasing one at large randomness with a minimum around a few $\bar{E}$. Furthermore, the $\theta$-dependence of $\xi$ in both regions shows a $\xi \propto \sin^{-2}\theta$ behavior as shown in Fig. 3a. The difference in the $\theta$-dependence of $\xi$ in the two pseudospin systems is due to different scattering potentials for oblique waves. In the following, we present analytical derivations of the localization length for both systems.

**Transformation from a Vector Wave Equation to a Scalar One.** For the layered structure, the wavevector component parallel to the interface ($k_y = k_0 \sin\theta$ where $k_0 = \bar{E}$ is the wavevector in the background) is conserved, with the same $k_y$ value in all layers. Thus, the wave functions for pseudospin-1 systems can be written as $\psi(x,y) = (\psi_1(x), \psi_2(x), \psi_3(x))^T e^{ik_y y}$. By using the following matrix representation for the spin operator, $\vec{\mathrm{S}} = S_x\hat{x} + S_y\hat{y}$:

$$S_x = \frac{1}{\sqrt{2}}\begin{pmatrix} 0 & 1 & 0 \\ 1 & 0 & 1 \\ 0 & 1 & 0 \end{pmatrix}, \quad S_y = \frac{1}{\sqrt{2}}\begin{pmatrix} 0 & -i & 0 \\ i & 0 & -i \\ 0 & i & 0 \end{pmatrix}, \tag{4}$$

we rewrite Eq. (2) as

$$\frac{1}{\sqrt{2}}\begin{pmatrix} 0 & -i\frac{\partial}{\partial x}-ik_y & 0 \\ -i\frac{\partial}{\partial x}+ik_y & 0 & -i\frac{\partial}{\partial x}-ik_y \\ 0 & -i\frac{\partial}{\partial x}+ik_y & 0 \end{pmatrix}\begin{pmatrix}\psi_1\\ \psi_2\\ \psi_3\end{pmatrix}+\bar{V}(x)\begin{pmatrix}\psi_1\\ \psi_2\\ \psi_3\end{pmatrix}=\bar{E}\begin{pmatrix}\psi_1\\ \psi_2\\ \psi_3\end{pmatrix}. \tag{5}$$

By eliminating $\psi_1(x)$ and $\psi_3(x)$, we can convert Eq. (5) into a scalar wave equation for $\psi_2(x)$,

$$-\frac{d}{dx}\left(\frac{1}{\bar{E}-\bar{V}(x)}\frac{d\psi_2}{dx}\right)+\frac{k_y^2}{\bar{E}-\bar{V}(x)}\psi_2=\left(\bar{E}-\bar{V}(x)\right)\psi_2 . \tag{6}$$

Without loss of generality, we take the first interface of the *N*-layer system as the origin, define a new dimensionless coordinate variable $u\equiv\int_0^x(\bar{E}-\bar{V}(x))dx$, and write $\Psi(u)\equiv\psi_2(x)$ and $\bar{U}(u)\equiv\bar{V}(x)$. Then, Eq. (6) can be re-expressed as

$$\frac{d^2\Psi}{du^2}+\Psi=\frac{k_y^2}{\left(\bar{E}-\bar{U}(u)\right)^2}\Psi . \tag{7}$$

By using the above coordinate transformation, we have transformed a non-standard wave equation, Eq. (6), to a standard one, Eq. (7), where the scattering potential due to the disordered potential $V(x)$ is explicitly shown on the right hand side of Eq. (7). In the case of normal incidence, i.e., $k_y=0$, Eq. (7) describes wave propagation in a homogeneous medium and contains two general solutions $\Psi\propto e^{\pm iu}=\exp\left[\pm i\int_0^x(\bar{E}-\bar{V}(x))dx\right]$. Thus, the accumulated random phase due to $V(x)$ during the one-way transport is now absorbed in the new coordinate *u*. For the layered structure where the potential is piece-wise constant, the *i*-th interface in the *u* coordinate, $u_i$, is written as $u_1=0$ and $u_i=\sum_{j=1}^{i-1}(\bar{E}-\bar{\mathrm{v}}_j)d$ for $i\geq 2$ from the above coordinate transformation. It is important to point out that we have transformed a three-component vector wave equation for obliquely propagating waves, i.e. Eq. (1), into an equivalent scalar wave equation for normally propagating waves, and the oblique angle enters the wave equation in the scattering terms, i.e., Eq. (7). Such a transformation allows us to derive analytically certain asymptotic localization behaviors.

Similarly, we can use the Pauli matrices for the spin-1/2 operator in Eq. (1) to construct a scalar wave equation for pseudospin-1/2 systems. In the *u* coordinate system, the wave equation has the form (see Supporting Information),

$$\frac{d^2\Psi}{du^2}+\Psi=\frac{k_y^2}{\left(\bar{E}-\bar{U}(u)\right)^2}\Psi+k_y\Psi\sum_{i=1}^{N+1}U_i\delta(u-u_i)\,, \tag{8}$$

where $U_i=\frac{1}{\bar{E}-\bar{\mathrm{v}}_i}-\frac{1}{\bar{E}-\bar{\mathrm{v}}_{i-1}}$. Note that in comparison with pseudospin-1 systems, pseudospin-1/2 systems have additional interface scattering terms $k_y\Psi\sum_{i=1}^{N+1}U_i\delta(u-u_i)$ located at all *N*+1 interfaces.

The difference in the $\theta$-dependence of $\xi$ in the two systems shown in Fig. 3a, when $\bar{W}$ is small, can be qualitatively understood from the scattering terms in Eqs. (7) and (8). For ordinary disordered media, it is well accepted that the localization length in 1D systems is on the order of the mean free path, which is inversely proportional to the square of the scattering strength [3]. In the case of small $k_y$, the $k_y^2$ dependence in the effective scattering potential of Eq. (7) gives rise to a $k_y^{-4}$ (or $\sin^{-4}\theta$) behavior in the localization length, whereas the $k_y$ dependence in the interface scattering terms of Eq. (8) dominates and leads to a $k_y^{-2}$ (or $\sin^{-2}\theta$) behavior. The sudden drop of localization length near $\bar{W}=\bar{E}$ for pseudospin-1 systems is due to the diverging scattering term in Eq. (7) when $|\bar{E}-\bar{U}(u)|<|k_y|$ in some layers so that the waves become evanescent inside those layers. We will show analytically that it is the existence of those evanescent waves that changes the $\theta$-dependence of $\xi$ from $\xi\propto\sin^{-4}\theta$ in the region $\bar{W}<\bar{E}$ to $\xi\propto\sin^{-2}\theta$ in the region $\bar{W}>\bar{E}$. When $\bar{W}$ goes beyond its critical value $\bar{E}$, the probability of having evanescent waves is reduced with increasing $\bar{W}$ , and in the meantime, the scattering potentials in the propagating layers are weakened in general. As a result, $\xi$ increases with $\bar{W}$ . However, such a sudden drop of $\xi$ is smeared out by the interface scattering terms in Eq. (8) so that a smooth change of localization behaviors is found for pseudospin-1/2 systems.

**Lyapunov Exponent Obtained by the Surface Green Function (SGF) Method.** Since Eqs. (7) and (8) are already in the form of scalar wave equations for normally propagating waves, we can now solve the wave localization problems of pseudospin systems by using the SGF method proposed for random layered systems [50]. The SGF method gives the following expression for the transmission coefficient of a normally incident plane wave propagating through a *N*- layered random system [50]:

$$T_N = |D_{N+1}|^{-2}, \tag{9}$$

where

$$\frac{D_{N+1}}{D_{N+1}^0} = \left[ e^{2i\Phi_{1,N+1}} \prod_{n=1}^{N+1} (1 - r_{n,n-1})(1 - r_{n-1,n}) \right]^{-1/2}. \tag{10}$$

Here $r_{n,n-1}$ denotes the reflection amplitude of a plane wave incident from the *n*-th layer on the (*n*-1)-th layer, $\Phi_{i,j} = \Phi_{j,i}$ is the phase accumulation between the *i*-th and *j*-th interfaces of the sample, and $D_{N+1}^0$ is the determinant of a *N*+1 by *N*+1 matrix $\hat{D}_{N+1}^0$ with the following elements,

$$(\hat{D}_{N+1}^0)_{nk} = \begin{cases} \delta_{nk} + (1 - \delta_{nk}) r_{k,k-1} e^{i\Phi_{n,k}} & n \ge k, \\ \delta_{nk} + (1 - \delta_{nk}) r_{k-1,k} e^{i\Phi_{n,k}} & n \le k. \end{cases} \tag{11}$$

As shown in Supporting Information, for both pseudospin systems, the phase accumulation can be expressed as

$$\Phi_{i,j} = \sum_{n=i}^{j-1} \bar{k}_n (\bar{E} - \mathrm{v}_n) d, \qquad \bar{k}_n = \sqrt{1 - \frac{k_y^2}{(\bar{E} - \mathrm{v}_n)^2}} \ , \tag{12}$$

and the reflection amplitudes can be written as

$$r_{n-1,n} = -r_{n,n-1} = \frac{\bar{k}_n - \bar{k}_{n-1}}{\bar{k}_n + \bar{k}_{n-1}}, \tag{13}$$

for pseudospin-1 systems, and

$$r_{n-1,n} = \frac{\bar{k}_n - \bar{k}_{n-1} + ik_y U_n}{\bar{k}_n + \bar{k}_{n-1} + ik_y U_n}, \qquad r_{n,n-1} = -\frac{\bar{k}_n - \bar{k}_{n-1} - ik_y U_n}{\bar{k}_n + \bar{k}_{n-1} + ik_y U_n}, \tag{14}$$

for pseudospin-1/2 systems. From Eqs. (9) and (10), we obtain the expression for Lyapunov exponent $\gamma$ in Eq. (3) as

$$\gamma = \xi^{-1} = \gamma_1 + \gamma_2, \tag{15}$$

with $\gamma_1 \equiv \frac{1}{Nd}\left\langle \ln |D_{N+1}^0| \right\rangle_c$ and $\gamma_2 \equiv -\frac{1}{2Nd}\left\langle \ln \left| e^{2i\Phi_{1,N+1}} \prod_{n=1}^{N+1} (1-r_{n,n-1})(1-r_{n-1,n}) \right| \right\rangle_c$. We first numerically calculate the localization length by using Eq. (15) as a function of $\bar{W}$ for a fixed incident angle and energy. The results are shown by black dashed lines in Figs. 2a and 2c for pseudospin-1 and -½ systems, respectively. We can see that they are in excellent agreements with those obtained from the TMM method.

**Asymptotic $\theta$-dependent Localization Length Behavior in Region $\bar{W} < \bar{E}$.** In the following, using Eq. (15), we show analytically that localization length follows the asymptotic behavior $\xi \propto \sin^{-4}\theta$ in the region of $\bar{W} < \bar{E}$. In this case, the reflection amplitudes in pseudospin-1 systems can be approximated as $r_{n-1,n} = -r_{n,n-1} \approx -\frac{k_y^2}{4}\left[\frac{1}{(\bar{E}-\bar{\mathrm{v}}_n)^2} - \frac{1}{(\bar{E}-\bar{\mathrm{v}}_{n-1})^2}\right]$, as long as $|k_y| << |\bar{E} - \bar{U}(u)|$. In this limit, as shown in Supporting Information, the Lyapunov exponent $\gamma$ can be written as

$$\gamma = \xi^{-1} = \gamma_1 + \gamma_2 \approx \frac{\bar{E}^4 \sin^4\theta}{32d}\left(\alpha_1 + \alpha_2\right), \tag{16}$$

where $\alpha_1$ and $\alpha_2$ are coefficients corresponding to $\gamma_1$ and $\gamma_2$, respectively (see Supporting Information). Note that $\gamma$ in Eq. (16) is proportional to $\sin^4\theta$ for the region $\bar{W} < \bar{E}$. In the case of $\bar{W} << \bar{E}$, we can further take a small $\bar{\mathrm{v}}_n / \bar{E}$ expansion for $\alpha_1$ and $\alpha_2$. It can be shown (see Supporting Information) that $\alpha_1$ and $\alpha_2$ then reduce to simple forms, $\alpha_1 \approx -\frac{8\bar{W}^2}{3\bar{E}^6}\cos\left(2\bar{E}d\cos\theta\right)$ and $\alpha_2 \approx \frac{8\bar{W}^2}{3\bar{E}^6}$. Thus, Eq. (16) gives the following expression for $\gamma$ in the limit $\bar{W} << \bar{E}$:

$$\gamma = \xi^{-1} \approx \frac{\bar{W}^2\sin^4\theta}{12\bar{E}^2 d}\left[1-\cos\left(2\bar{E}d\cos\theta\right)\right]. \qquad (17)$$

We have also numerically calculated the localization length in this limit. The results are shown in Fig. 3b by the symbols. We find excellent agreement between the analytical and numerical results. It is interesting to note that the value of $\gamma$ vanishes at certain energies which satisfy the on-average Fabry-Perot resonance condition $\bar{E}d\cos\theta = m\pi$ ($m \in$ integers). Thus, $\xi$ tends to diverge at these energies. The finite values of $\xi$ at these resonances are due to high-order corrections.

For pseudospin-1/2 systems, the asymptotic behavior of $\gamma$ in the limit of small $k_y$ and $\bar{W} << \bar{E}$ can be obtained using a similar approach (see Supporting Information) and has the expression

$$\gamma = \xi^{-1} \approx \frac{\bar{W}^2\sin^2\theta}{12\bar{E}^2 d}\left[1-\cos\left(2\bar{E}d\cos\theta\right)\right]. \qquad (18)$$

The validity of Eq. (18) is also confirmed numerically (see Fig. S1 in Supporting Information). From Eqs. (17) and (18), we can see that in both pseudospin systems, the localization length decreases as $\xi \propto \bar{W}^{-2}$, showing exactly the same behaviors in Fig. 2. More importantly, our analytical results prove that the pseudospin number indeed makes a profound difference on the localization behaviors, leading to a $\xi \propto \sin^{-4S}\theta$ localization length behavior for small $\theta$, where *S* is the pseudospin number.

**Asymptotic $\theta$-dependent Localization Length Behavior in Region $\bar{W} > \bar{E}$.** In this case, there are strong scatterings for those layers with the potentials $\bar{\mathrm{v}}$ close to the incident energy $\bar{E}$ due to the existence of singularity at $\bar{E} = \bar{\mathrm{v}}$ in the scattering potential in Eq. (7), and hence the approximations used above are not applicable. Although the calculation becomes rather tedious, we can still manage to obtain an analytic form of $\gamma_2$ for pseudospin-1 systems (see Supporting Information), that is,

$$\gamma_2 \approx -\frac{1}{2d}\left\langle \ln\left|1 - r_{n,n-1}^2\right|\right\rangle_c \approx \frac{\bar{E}^2 \sin^2\theta}{2\bar{W}^2 d} \ . \qquad (19)$$

In order to confirm the validity of Eq. (19), we numerically calculate $\gamma_2$ as a function of the incident angle for $\bar{W} = 0.03$ and $\bar{E} = 0.02$. The result is plotted by red circles in Fig. 4a, which agrees excellently with the analytic expression (red solid line) shown in Eq. (19). Since the Lyapunov exponent, $\gamma = \gamma_1 + \gamma_2$, is an even function of $\sin\theta$, we can safely conclude from Eq. (19) that the region $\bar{W} > \bar{E}$ represents a different localization phase in which the $\theta$-dependent localization length has an asymptotic behavior, $\gamma = \xi^{-1} \propto \sin^2\theta$, different from the $\xi^{-1} \propto \sin^4\theta$ behavior found in the region $\bar{W} < \bar{E}$ as shown in Eq. (16). Such a sudden change of $\theta$-dependent localization behavior at $\bar{W} = \bar{E}$ is accompanied by the cusp-like change of localization length from a decreasing function of $\bar{W}$ when $\bar{W} < \bar{E}$ to an increasing one when $\bar{W} > \bar{E}$ as shown in Figs. 2a and 2b. We show in the Supporting Information that the origin of the $\sin^2\theta$ factor in $\gamma_2$ is the occurrence of the diverging scattering potentials in certain layers when $|k_y| > |\bar{E} - \bar{U}(u)|$ so that the waves become evanescent inside these layers. In fact, the presence of evanescent waves in certain layers also leads to a $\sin^2\theta$ dependence in $\gamma_1$. Due to the complexity of the matrix $\hat{D}_{N+1}^0$, an explicit analytic expression for $\gamma_1$ is formidable. We numerically calculate $\gamma_1$ and plot the result by green triangles in Fig. 4a, which has an excellent fit to a dotted line showing $\gamma_1 \propto \sin^2\theta$. If the presence of evanescent waves is the origin which turns a $\sin^4\theta$ dependence of $\gamma$ into a $\sin^2\theta$ dependence, we should be able to recover the $\sin^4\theta$ behavior found in region $\bar{W} < \bar{E}$ by purposely excluding evanescent waves in the random media. In order to confirm this point, we calculate the $\theta$-dependence of $\gamma$ for a particular random distribution of potentials, $\bar{\mathrm{v}} \in [-\bar{W}, \bar{W}]$ but with a condition $|\bar{E} - \bar{\mathrm{v}}| > 0.1\bar{E}$ so that no evanescent waves will occur at sufficiently small $\theta$. The result is plotted by blue squares in Fig. 4a. It is clearly seen that the $\gamma \propto \sin^4\theta$ behavior is indeed recovered. In fact, the sudden drop of $\xi$ near $\bar{W}_c = \bar{E}$ shown in Figs. 2a and 2b is also due to the presence of evanescent waves in some layers. To show this, we numerically calculate $\xi$ as a function of $\bar{W}$ by excluding the evanescent waves. The result is plotted by a blue dashed line in Fig. 4b. In comparison with the result with evanescent waves included (blue circles), we can see that the sudden drop of $\xi$ near $\bar{W}_c = \bar{E}$ disappears.

However, for pseudospin-1/2 systems, propagating waves also contribute to $\gamma$ a $\sin^2\theta$ term due to the interface scattering terms in Eq. (8), which smears out the sudden drop of $\xi$, as shown in Figs. 2c and 2d , and leads to the same asymptotic $\theta$-dependence of $\xi$ for all $\bar{W}$s in Fig. 3a.

## Conclusions

We discovered interesting anomalous Anderson localization behaviors in disordered pseudospin-1 systems employing numerical results by the TMM method as well as analytical solutions from the SGF method. In contrast to ordinary 1D random media where stronger randomness always induces stronger localization, pseudospin-1 systems have a critical random strength $\bar{W}_c = \bar{E}$ at which a cusp-like turnaround occurs in the localization length as a function of randomness. Additional randomness beyond this critical strength makes the wave less localized. Such a sudden change gives rise to two localization phases as characterized by different asymptotic $\theta$-dependence of the localization length, i.e., $\xi \propto \sin^{-4}\theta$ when $\bar{W} < \bar{W}_c$ and $\xi \propto \sin^{-2}\theta$ when $\bar{W} > \bar{W}_c$. Such anomalous behaviors arise from the existence of a Dirac-like point and the occurrence of the evanescent waves in the region $\bar{W} > \bar{W}_c$. For pseudospin-1/2 systems, we find that the sharp transition is smeared out by additional interface scattering terms and the localization length behavior shows a smooth change from decreasing with the random strength at small $\bar{W}$ to increasing at large $\bar{W}$. In both regions, the $\theta$-dependence of $\xi$ follows the same asymptotic behavior $\xi \propto \sin^{-2}\theta$. Recently pseudospin-1 systems have been experimentally realized in photonic [29, 30, 38-41] and ultracold atom systems [37]. Meanwhile, the applied potentials in such systems, as an analogy of gate voltage in graphene, can be realized by uniformly scaling the structure in PCs [31] or manipulating appropriate holographic mask in ultracold systems [32-36]. Thus, it is experimentally feasible to prepare a 1D disordered pseudospin-1 system using such artificial structures. For a given randomness $W$, two localization phases can be observed by tuning the incident energy from $E > W$ to $E < W$.

## Acknowledgements

This work was supported by a grant from the Research Grants Council of the Hong Kong (Project No. AoE/P-02/12). S.G.L. also acknowledges partial support from the National Science Foundation under grant DMR-1508412.

# Figure legends

Fig. 1. Schematic diagram of 1D disordered systems. (a) Top view of the structure. Each layer has the same thickness *d*, but feels a randomized potential. (b) One possible realization of random potentials. The potentials V are uniformly distributed in the range $[-W,\ W]$.

Fig. 2. Localization length as a function of normalized random potential strength for different incident angles and energies in 1D disordered pseudospin-1 and -1/2 systems calculated with the transfer-matrix method (TMM). (a) Localization length for three different incident angles in pseudospin-1 systems. (b) Localization length for three different incident energies in pseudospin-1 systems. (c) Same as (a), but for pseudospin-½ systems. (d) Same as (b), but for pseudospin-½ systems. The black dash lines in (a) and (c) show the results obtained by the surface Green function (SGF) method. The localization lengths for small $\bar{W}$ are fitted by dotted lines, showing an asymptotic behavior $\xi \propto \bar{W}^{-2}$.

Fig. 3. Localization behaviors for disordered pseudospin-1 and -1/2 systems. (a) Localization length as a function of incident angle for incident energy $\bar{E} = 0.02$ and two random strengths in 1D disordered pseudospin- 1 and -1/2 systems calculated using the transfer-matrix method (TMM). The two random strengths are chosen from the respective decreasing and increasing regions in Figs. 2a and 2c for pseudospin-1 and -1/2 systems. The localization length of pseudospin-1 systems at small $\theta$ for $\bar{W} = 0.01$ ($< \bar{E}$) (blue circles) is fitted by a dotted line, showing $\xi \propto \sin^{-4}\theta$. All other three cases are fitted by $\xi \propto \sin^{-2}\theta$. (b) Comparison of the localization length calculated by using the TMM method and analytical results shown in Eq. (17). Both $\bar{E}$ and $\bar{W}$ are in unit of $2\pi / d$.

Fig. 4. Effect of evanescent waves on the localization behaviors in pseudospin-1 systems. (a) Comparison of the Lyapunov exponents as a function of incident angle with and without evanescent waves included in 1D disordered pseudospin-1 systems at $\bar{E} = 0.02$ and $\bar{W} = 0.03$. In the case with evanescent waves, $\gamma_1$ at small $\theta$ (green triangles) is fitted by a dotted line showing $\gamma_1 \propto \sin^2\theta$, and the numerical result of $\gamma_2$ (red circles) agrees excellently with the analytic prediction in Eq. (19) (red solid line). For the random distribution of potentials $|\bar{E} - \bar{\mathrm{v}}| > 0.1\bar{E}$, no evanescent waves occur at sufficiently small $\theta$. $\gamma$ in this case (blue squares) shows an excellent fit to a dotted line $\gamma \propto \sin^4\theta$ for small $\theta$. (b) Comparison between the localization lengths with and without evanescent waves for pseudospin-1 systems with $\bar{E} = 0.02$ and $\sin\theta = 0.3$. Both $\bar{E}$ and $\bar{W}$ are in unit of $2\pi / d$.

## Figures

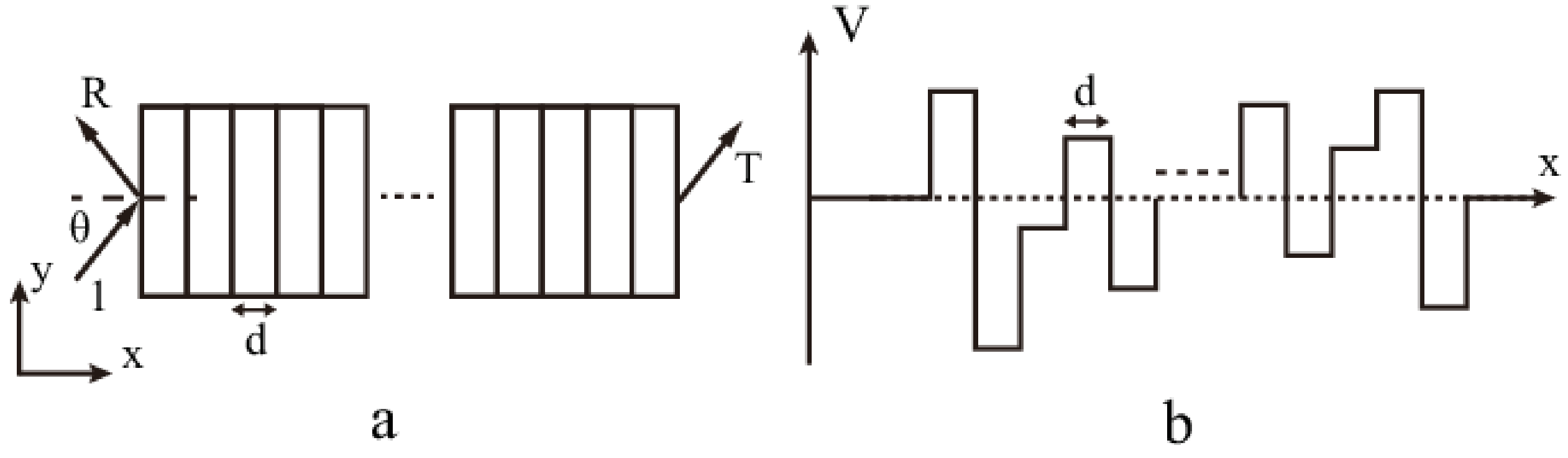

Figure 1

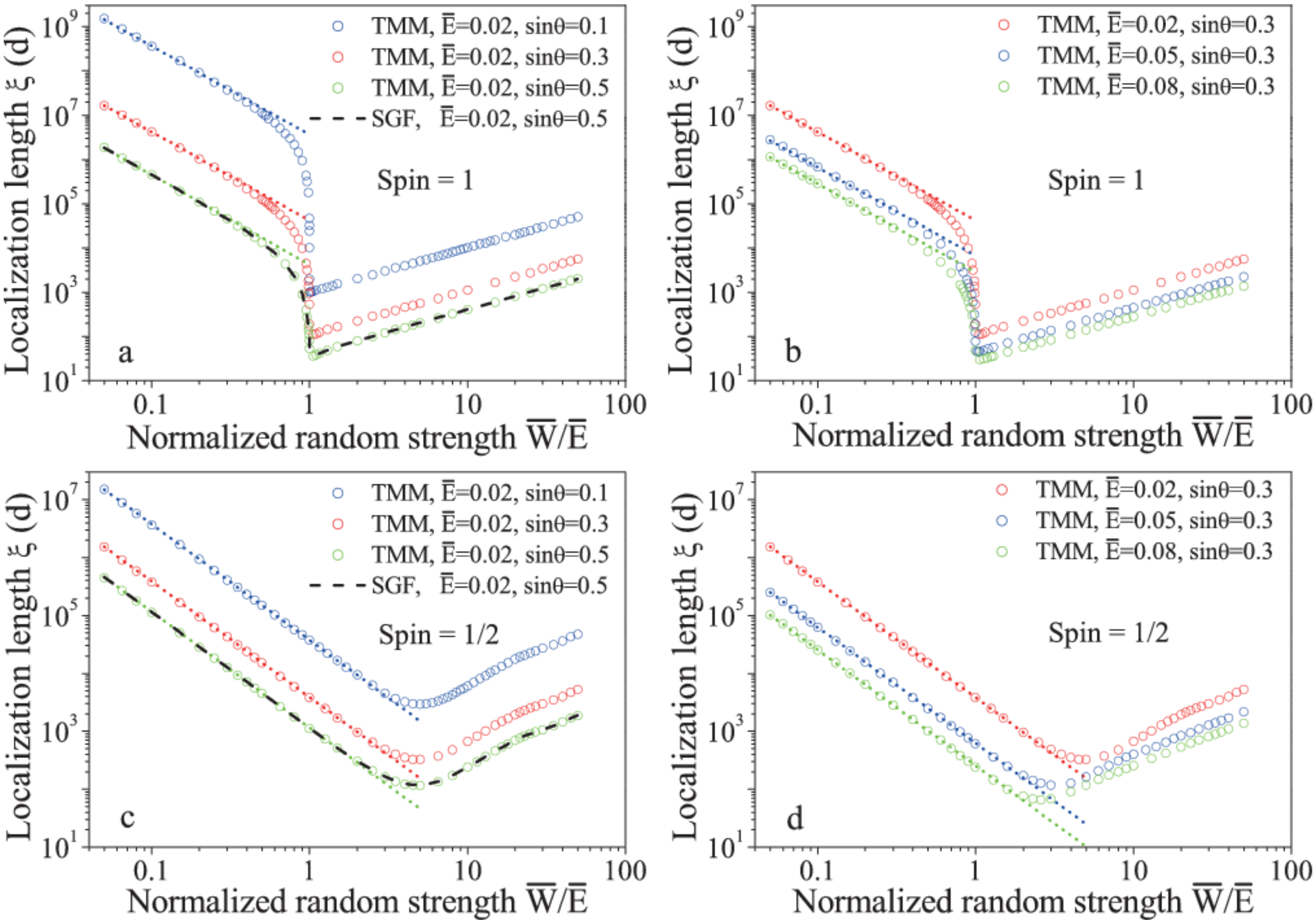

Figure 2

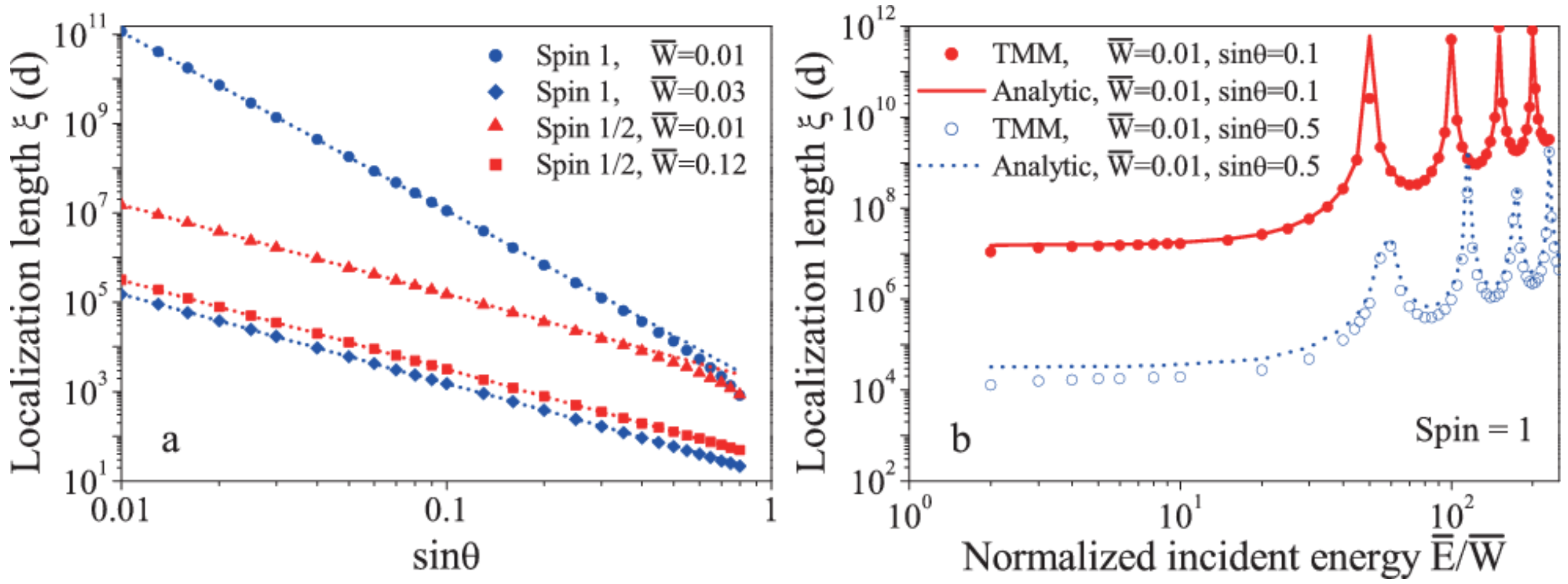

Figure 3

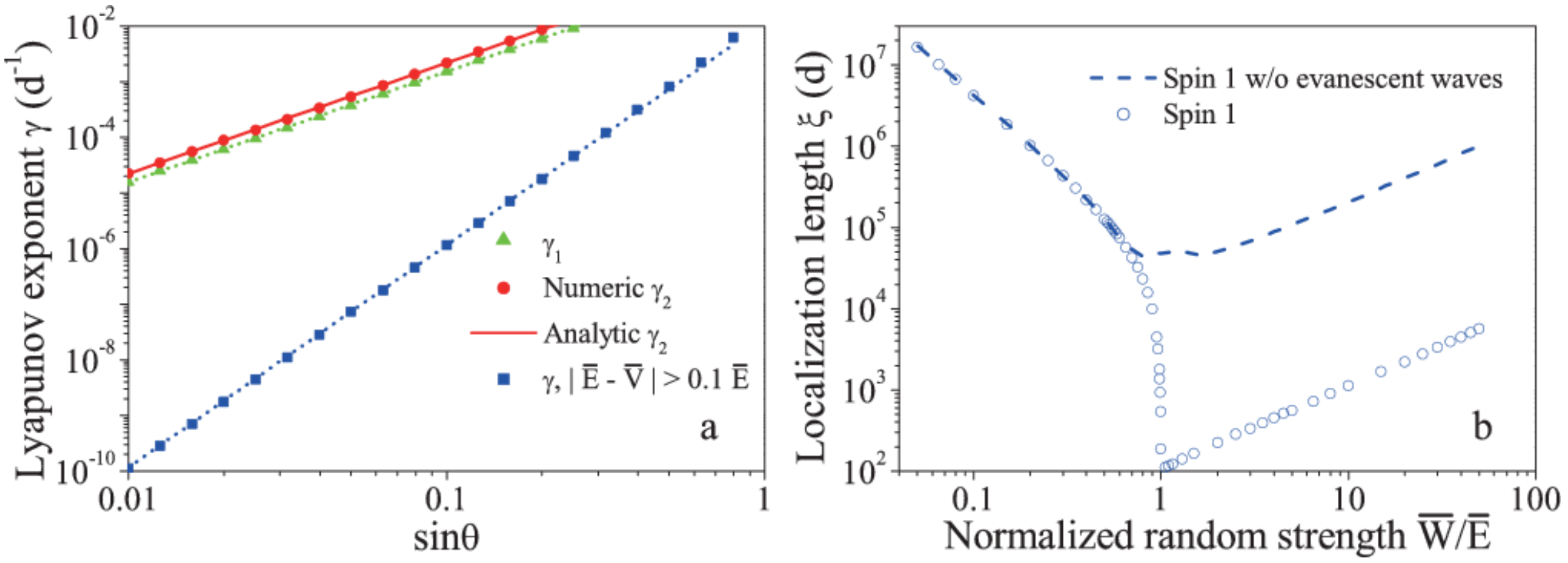

Lyapunov exponent γ (d⁻¹)
sinθ
γ₁
Numeric γ₂
Analytic γ₂
γ, | Ē - V̄ | > 0.1 Ē
a
Localization length ξ (d)
Normalized random strength W̄/Ē
Spin 1 w/o evanescent waves
Spin 1
b

Figure 4

# Supporting Information

### Derivation of the Scalar Wave Equation for Pseudospin-1/2 Systems

For pseudospin-1/2 systems, $\vec{\mathrm{S}}$ in Eq. (2) is a 2D Pauli vector, i.e., $\vec{\mathrm{S}} = \sigma_x \hat{x} + \sigma_y \hat{y}$ with

$$\sigma_x = \begin{pmatrix} 0 & 1 \\ 1 & 0 \end{pmatrix}, \quad \sigma_y = \begin{pmatrix} 0 & -i \\ i & 0 \end{pmatrix}, \tag{S1}$$

and the wave function $\psi$ is a two-component spinor function , $\psi = (\psi_1(x), \psi_2(x))^T e^{ik_y y}$ , since for oblique incidence on the layered structure, the wavevector component parallel to the interface, $k_y$ , is conserved in each layer. Taking the operator $\vec{\mathrm{k}} = k_x \hat{x} + k_y \hat{y}$ as $k_x = -i\frac{\partial}{\partial x}$ and $k_y = -i\frac{\partial}{\partial y}$ , we can express Eq. (2) for pseudospin-1/2 systems as,

$$\begin{pmatrix} 0 & -i\frac{\partial}{\partial x} - ik_y \\ -i\frac{\partial}{\partial x} + ik_y & 0 \end{pmatrix} \begin{pmatrix} \psi_1 \\ \psi_2 \end{pmatrix} + \bar{V}(x) \begin{pmatrix} \psi_1 \\ \psi_2 \end{pmatrix} = \bar{E} \begin{pmatrix} \psi_1 \\ \psi_2 \end{pmatrix}, \tag{S2}$$

or,

$$-i\frac{\partial \psi_2}{\partial x} - ik_y \psi_2 + \bar{V}(x)\psi_1 = \bar{E}\psi_1, \tag{S3}$$

$$-i\frac{\partial \psi_1}{\partial x} + ik_y \psi_1 + \bar{V}(x)\psi_2 = \bar{E}\psi_2. \tag{S4}$$

By eliminating $\psi_2$ in Eqs. (S3) and (S4), we obtain the following scalar wave equation for $\psi_1$:

$$\frac{-1}{\bar{E} - \bar{V}(x)} \frac{d}{dx}\left[\frac{1}{\bar{E} - \bar{V}(x)} \frac{d\psi_1}{dx}\right] + \frac{k_y \psi_1}{\bar{E} - \bar{V}(x)} \frac{d}{dx}\left(\frac{1}{\bar{E} - \bar{V}(x)}\right) + \frac{k_y^2}{\left(\bar{E} - \bar{V}(x)\right)^2}\psi_1 = \psi_1. \tag{S5}$$

We now define a new coordinate variable in the same way as that in pseudospin-1 systems, i.e., $u \equiv \int_0^x (\bar{E} - \bar{V}(x))dx$ , and write $\Psi(u) \equiv \psi_1(x)$ and $\bar{U}(u) \equiv \bar{V}(x)$ . Then, we can rewrite Eq. (S5) as

$$\frac{d^2\Psi}{du^2}+\Psi=\frac{k_y^2}{\left(\bar{E}-\bar{U}(u)\right)^2}\Psi+k_y\Psi\sum_{i=1}^{N+1}U_i\delta(u-u_i)\,, \tag{S6}$$

where $u_i$ is the *i*-th interface in the *u* coordinate with $u_1=0$ and $u_i=\sum_{j=1}^{i-1}(\bar{E}-\bar{\mathrm{v}}_j)d$ for $i\geq 2$, and $U_i=\frac{1}{\bar{E}-\bar{\mathrm{v}}_i}-\frac{1}{\bar{E}-\bar{\mathrm{v}}_{i-1}}$. Note that here in the *u* coordinate $\bar{E}-\bar{U}(u)$ is a piece-wise constant function since the normalized potential $\bar{U}(u)$ is a constant in each layer. Thus, we have $\frac{d}{du}\left(\frac{1}{\bar{E}-\bar{U}(u)}\right)=\sum_{i=1}^{N+1}U_i\delta(u-u_i)$. In comparison with Eq. (7), the scalar wave equation, Eq. (S6), for pseudospin-1/2 systems has additional interface scattering terms located at all $N+1$ interfaces.

**Phase Accumulation between Two Interfaces and Reflection Amplitudes at an Interface**

Inside the *n*-th layer, Eqs. (7) and (8) can be simplified as

$$\frac{d^2\Psi}{du^2}+\bar{k}_n^2\Psi=0\,, \tag{S7}$$

with $\bar{k}_n^2=1-\frac{k_y^2}{(\bar{E}-\bar{\mathrm{v}}_n)^2}$, for both pseudospin systems, where $\bar{\mathrm{v}}_n$ is the normalized potential in the *n*-th layer. Eq. (S7) is a wave equation in the *u* coordinate with the wave number $\bar{k}_n$. Thus, the accumulated phase between *i*-th and *j*-th interfaces of the sample is $\Phi_{i,j}=\Phi_{j,i}=\sum_{n=i}^{j-1}\bar{k}_n(\bar{E}-\bar{\mathrm{v}}_n)d$ ($i<j$), where $(\bar{E}-\bar{\mathrm{v}}_n)d$ is the thickness of *n*-th layer in the *u* coordinate, derived from the coordinate transformation $u\equiv\int_0^x(\bar{E}-\bar{V}(x))dx$.

Now we derive the reflection amplitudes between the ($n$-1)-th and $n$-th layers. Consider two semi-infinite homogeneous media meeting at an interface at $u = u_n$ : on the left of the interface ($u < u_n$) is the normalized potential $\overline{\mathrm{v}}_{n-1}$ in the ($n$-1)-th layer , and on the right ($u > u_n$) is $\overline{\mathrm{v}}_n$ in the $n$-th layer. Suppose that $G_m^{(0)}(u,u')$ ($m = n-1, n$) is the 1D Green's function for each medium when it is infinite, then we can construct the Green's function $G_m^{(1)}(u,u')$ ($m = n-1, n$) in each semi-infinite media in the presence of one interface [50] as follows,

$$
\begin{aligned}
G_{n-1}^{(1)}(u,u') &= G_{n-1}^{(0)}(u,u') - r_{n-1,n}\frac{G_{n-1}^{(0)}(u,u_n)G_{n-1}^{(0)}(u_n,u')}{G_{n-1}^{(0)}(u_n,u_n)} \qquad u,u' \le u_n, \\
G_{n}^{(1)}(u,u') &= G_{n}^{(0)}(u,u') - r_{n,n-1}\frac{G_{n}^{(0)}(u,u_n)G_{n}^{(0)}(u_n,u')}{G_{n}^{(0)}(u_n,u_n)} \qquad u,u' \ge u_n.
\end{aligned} \tag{S8}
$$

For pseudospin-1 systems, the Green's functions should satisfy the following boundary conditions at the interface $u = u_n$ obtained from Eq. (7),

$$G_{n-1}^{(1)}(u_n,u_n) = G_n^{(1)}(u_n,u_n), \tag{S9}$$

$$\dot{G}_n^{(1)}(u_n+0,u_n) - \dot{G}_{n-1}^{(1)}(u_n-0,u_n) = 1. \tag{S10}$$

Here the dot over $G$ denotes the derivative with respect to the first argument. Solving Eqs. (S9) and (S10) , we obtain

$$r_{n-1,n} = -r_{n,n-1} = \frac{G_{n-1}^{(0)} - G_n^{(0)}}{G_{n-1}^{(0)} + G_n^{(0)}}, \qquad (G_m^{(0)} \equiv G_m^{(0)}(u_n,u_n),\ m = n-1, n) \tag{S11}$$

for pseudospin-1 systems. Note that we have $\dot{G}_m^{(0)}(u \pm 0, u) = \pm\frac{1}{2}$ ($m = n-1, n$) for the homogeneous medium.

For pseudospin-1/2 systems, the Green's functions obey different boundary conditions due to the interface scattering potentials in Eq. (8),

$$G_{n-1}^{(1)}(u_n,u_n)=G_n^{(1)}(u_n,u_n)\,, \tag{S12}$$

$$\dot{G}_n^{(1)}(u_n+0,u_n)-\dot{G}_{n-1}^{(1)}(u_n-0,u_n)=1+k_yU_nG_{n-1}^{(1)}(u_n,u_n)\,, \tag{S13}$$

where $U_n=\dfrac{1}{\bar{E}-\bar{\mathrm{v}}_n}-\dfrac{1}{\bar{E}-\bar{\mathrm{v}}_{n-1}}$. Thus, we can obtain the reflection amplitudes for pseudospin-1/2 systems,

$$r_{n-1,n}=\frac{G_{n-1}^{(0)}-(2k_yU_nG_{n-1}^{(0)}+1)G_n^{(0)}}{G_{n-1}^{(0)}-(2k_yU_nG_{n-1}^{(0)}-1)G_n^{(0)}}\ , \tag{S14}$$

$$r_{n,n-1}=-\frac{G_{n-1}^{(0)}+(2k_yU_nG_{n-1}^{(0)}-1)G_n^{(0)}}{G_{n-1}^{(0)}-(2k_yU_nG_{n-1}^{(0)}-1)G_n^{(0)}}\,. \tag{S15}$$

For both systems, the Green's function in the medium of the *m*-th layer, $G_m^{(0)}(u,u')$, satisfies the following equation:

$$\frac{d^2G_m^{(0)}(u,u')}{du^2}+\bar{k}_m^2G_m^{(0)}(u,u')=\delta(u-u') \qquad \left(\bar{k}_m^2=1-\frac{k_y^2}{(\bar{E}-\bar{\mathrm{v}}_m)^2}\right). \tag{S16}$$

By solving Eq. (S16), we can obtain $G_m^{(0)}(u,u')=\dfrac{e^{i\bar{k}_m|u-u'|}}{2i\bar{k}_m}$. Substituting $G_m^{(0)}(u,u')$ into Eqs. (S11), (S14) and (S15), we have

$$r_{n-1,n}=-r_{n,n-1}=\frac{\bar{k}_n-\bar{k}_{n-1}}{\bar{k}_n+\bar{k}_{n-1}} \tag{S17}$$

for pseudospin-1 systems, and

$$r_{n-1,n}=\frac{\bar{k}_n-\bar{k}_{n-1}+ik_yU_n}{\bar{k}_n+\bar{k}_{n-1}+ik_yU_n}\,, \tag{S18}$$

$$r_{n,n-1}=-\frac{\bar{k}_n-\bar{k}_{n-1}-ik_yU_n}{\bar{k}_n+\bar{k}_{n-1}+ik_yU_n}\,, \tag{S19}$$

for pseudospin-1/2 systems.

Note that the reflection amplitudes obtained here from surface Green function method are consistent with the results in Ref. 31 calculated in *x*-coordinate by matching the boundary conditions.

**Derivation of the Lyapunov Exponent for Pseudospin-1 and -½ Systems in the Region of $\bar{W} < \bar{E}$**

In region $\bar{W} < \bar{E}$, the reflection amplitudes for pseudospin-1 systems can be approximated as

$r_{n-1,n} = -r_{n,n-1} \approx -\frac{k_y^2}{4}\left[\frac{1}{(\bar{E}-\bar{\mathrm{v}}_n)^2} - \frac{1}{(\bar{E}-\bar{\mathrm{v}}_{n-1})^2}\right]$ when $|k_y| << \bar{E} - \bar{U}(u)$. Note that

$\bar{k}_n = \sqrt{1 - \frac{k_y^2}{(\bar{E}-\bar{\mathrm{v}}_n)^2}}$ is real for small $k_y$ satisfying $\bar{E} - \bar{\mathrm{v}}_n > |k_y|$, then we have $|e^{2i\Phi_{1,N+1}}| = 1$ where

$\Phi_{1,N+1} = \sum_{n=1}^{N} \bar{k}_n (\bar{E} - \bar{\mathrm{v}}_n) d$. This indicates that $\Phi_{1,N+1}$ does not contribute to $\gamma_2$ shown in Eq. (15). Thus,

in the limit of $|k_y| << \bar{E} - \bar{U}(u)$, $\gamma_2$ can be approximated as

$$
\begin{aligned}
\gamma_2 &\approx -\frac{1}{2d}\left\langle \ln(1 - r_{n,n-1}^2) \right\rangle_c \approx \frac{1}{2d}\left\langle r_{n,n-1}^2 \right\rangle_c \\
&\approx \frac{k_y^4}{32d}\left\langle \left[\frac{1}{(\bar{E}-\bar{\mathrm{v}}_n)^2} - \frac{1}{(\bar{E}-\bar{\mathrm{v}}_{n-1})^2}\right]^2 \right\rangle_c \\
&= \frac{k_y^4}{32d}\alpha_2,
\end{aligned}
\tag{S20}
$$

where $\alpha_2 \equiv \left\langle \left[\frac{1}{(\bar{E}-\bar{\mathrm{v}}_n)^2} - \frac{1}{(\bar{E}-\bar{\mathrm{v}}_{n-1})^2}\right]^2 \right\rangle_c$.

The determinant $D_{N+1}^0$ can be calculated using the Leibniz formula. The Leibniz formula for the determinant of a $n \times n$ matrix *A* is

$$\det(A) = \sum (-1)^k a_{1k_1} a_{2k_2} \cdots a_{nk_n} \ . \tag{S21}$$

Here the sequence $[k_1, k_2, \cdots, k_n]$ is one permutation of the set $\{1, 2, \cdots, n\}$ achieved by successively interchanging two entries $k$ times. For a matrix with all diagonal elements as 1, Eq. (S21) can be written as a summation over a series of products with different number of off-diagonal elements,

$$\det(A) = 1 - \sum_{i<j} a_{ij} a_{ji} + \sum_{i \neq j \neq k,\, i \neq k} a_{ij} a_{jk} a_{ki} + \cdots \ , \tag{S22}$$

where the first term is the product of all diagonal elements for no interchange of the entries, the second term is the product of two off-diagonal elements for interchanging two entries once, and the third term is the product of three off-diagonal elements, etc. So for the matrix $\hat{D}^0_{N+1}$ with the elements in Eq. (11) and $r_{n-1,n} = -r_{n,n-1} \approx -\frac{k_y^2}{4}\left[\frac{1}{(\bar{E}-\bar{\mathrm{v}}_n)^2} - \frac{1}{(\bar{E}-\bar{\mathrm{v}}_{n-1})^2}\right]$, the determinant $D^0_{N+1}$ can be calculated to the lowest order of $k_y$,

$$D^0_{N+1} = 1 + \frac{k_y^4}{16} \sum_{i<j} \left[\frac{1}{(\bar{E}-\bar{\mathrm{v}}_i)^2} - \frac{1}{(\bar{E}-\bar{\mathrm{v}}_{i-1})^2}\right]\left[\frac{1}{(\bar{E}-\bar{\mathrm{v}}_j)^2} - \frac{1}{(\bar{E}-\bar{\mathrm{v}}_{j-1})^2}\right] e^{2i\Phi_{i,j}} + O(k_y^6) \ . \tag{S23}$$

Here the phase factor $e^{2i\Phi_{i,j}} = \exp\left[2i\sum_{n=i}^{j-1} \bar{k}_n (\bar{E}-\bar{\mathrm{v}}_n) d\right]$ with $\bar{k}_n = \sqrt{1 - \frac{k_y^2}{(\bar{E}-\bar{\mathrm{v}}_n)^2}}$ . So, for small $k_y$ the determinant $D^0_{N+1}$ can be written as

$$D^0_{N+1} \approx 1 + N \frac{k_y^4}{16} \kappa(\bar{E}, k_y) \ , \tag{S24}$$

where $\kappa(\bar{E}, k_y) \approx \frac{1}{N} \sum_{i<j} \left\langle \left[\frac{1}{(\bar{E}-\bar{\mathrm{v}}_i)^2} - \frac{1}{(\bar{E}-\bar{\mathrm{v}}_{i-1})^2}\right]\left[\frac{1}{(\bar{E}-\bar{\mathrm{v}}_j)^2} - \frac{1}{(\bar{E}-\bar{\mathrm{v}}_{j-1})^2}\right] e^{2i\Phi_{i,j}} \right\rangle_c$ . Thus, we can obtain the following asymptotic behavior of $\gamma_1$ in the limit of $|k_y| << \bar{E} - \bar{U}(u)$ ,

$$\gamma_1 = \frac{1}{Nd}\left\langle \ln |D^0_{N+1}| \right\rangle_c \approx \alpha_1 \frac{k_y^4}{32d}, \tag{S25}$$

where $\alpha_1 = 2\Re(\kappa)$ and $\Re(\kappa)$ denotes the real part of $\kappa$. Then the Lyapunov exponent for the case $\bar{W} < \bar{E}$ can be obtained as

$$\gamma = \xi^{-1} = \gamma_1 + \gamma_2 \approx \frac{k_y^4}{32d}\left(\alpha_1 + \alpha_2\right). \tag{S26}$$

Note that the Lyapunov exponent $\gamma$ is proportional to $k_y^4$ (or $\sin^4\theta$ ) in Eq. (S26).

In the case of $\bar{W} << \bar{E}$, we can further take a small $\bar{\mathrm{v}}_n / \bar{E}$ expansion in the expression of $\alpha_2$, and find

$$\alpha_2 \equiv \left\langle \left[ \frac{1}{(\bar{E}-\bar{\mathrm{v}}_n)^2} - \frac{1}{(\bar{E}-\bar{\mathrm{v}}_{n-1})^2} \right]^2 \right\rangle_c \approx \frac{4}{\bar{E}^6}\left\langle \left(\bar{\mathrm{v}}_n - \bar{\mathrm{v}}_{n-1}\right)^2 \right\rangle_c = \frac{8\bar{W}^2}{3\bar{E}^6}. \tag{S27}$$

Here the ensemble averages of $\bar{\mathrm{v}}_{n-1}^2$, $\bar{\mathrm{v}}_n^2$, $\bar{\mathrm{v}}_{n-1}$ and $\bar{\mathrm{v}}_n$ are taken as

$\left\langle \bar{\mathrm{v}}_n^2 \right\rangle_c = \left\langle \bar{\mathrm{v}}_{n-1}^2 \right\rangle_c = \frac{1}{2W}\int_{-\bar{W}}^{\bar{W}} \bar{\mathrm{v}}^2 d\bar{\mathrm{v}} = \frac{1}{3}\bar{W}^2$ and $\left\langle \bar{\mathrm{v}}_{n-1} \right\rangle_c = \left\langle \bar{\mathrm{v}}_n \right\rangle_c = 0$ since $\bar{\mathrm{v}}_{n-1}$ and $\bar{\mathrm{v}}_n$ are independent random variables distributed uniformly in the range of $[-\bar{W}, \bar{W}]$. With the same approximation, $\kappa(\bar{E}, k_y)$ can be written as

$$\begin{aligned}\kappa(\bar{E}, k_y) &\approx \frac{1}{N}\sum_{i<j}\left\langle \left[ \frac{1}{(\bar{E}-\bar{\mathrm{v}}_i)^2} - \frac{1}{(\bar{E}-\bar{\mathrm{v}}_{i-1})^2} \right]\left[ \frac{1}{(\bar{E}-\bar{\mathrm{v}}_j)^2} - \frac{1}{(\bar{E}-\bar{\mathrm{v}}_{j-1})^2} \right] e^{2i\Phi_{i,j}} \right\rangle_c \\ &\approx \frac{4}{N\bar{E}^6}\sum_{i<j}\left\langle \left(\bar{\mathrm{v}}_i - \bar{\mathrm{v}}_{i-1}\right)\left(\bar{\mathrm{v}}_j - \bar{\mathrm{v}}_{j-1}\right) e^{2i\Phi_{i,j}} \right\rangle_c .\end{aligned} \tag{S28}$$

In the case of $\bar{\mathrm{v}} << \bar{E}$, $\bar{k}_n = \sqrt{1-\frac{k_y^2}{(\bar{E}-\bar{\mathrm{v}}_n)^2}} \approx \sqrt{1-\frac{k_y^2}{\bar{E}^2}\left(1+\frac{2\bar{\mathrm{v}}_n}{\bar{E}}\right)} \approx \sqrt{1-\frac{k_y^2}{\bar{E}^2}}\left(1-\frac{\bar{\mathrm{v}}_n k_y^2}{\bar{E}(\bar{E}^2-k_y^2)}\right)$. Thus,

in this limit, for the phase factor $\Phi_{i,j}$, we can obtain $\bar{k}_n(\bar{E}-\bar{\mathrm{v}}_n)d \approx k_{x0}d - \left(k_{x0}+\frac{k_y^2}{k_{x0}}\right)d\,\frac{\bar{\mathrm{v}}_n}{\bar{E}}$ with

$k_{x0} = \sqrt{\bar{E}^2-k_y^2}$. Thus, for $j > i+1$,

$$
\begin{aligned}
e^{2i\Phi_{i,j}} &= \exp\left\{2i\left[\bar{k}_i(\bar{E}-\bar{\mathrm{v}}_i)d + \bar{k}_{j-1}(\bar{E}-\bar{\mathrm{v}}_{j-1})d\right]\right\}\cdot\exp\left[2i\sum_{n=i+1}^{j-2}\bar{k}_n(\bar{E}-\bar{\mathrm{v}}_n)d\right] \\
&\approx e^{4ik_{x0}d}\left[1-2i\left(k_{x0}+\frac{k_y^2}{k_{x0}}\right)d\,\frac{\bar{\mathrm{v}}_i+\bar{\mathrm{v}}_{j-1}}{\bar{E}}\right]\cdot\exp\left[2i\sum_{n=i+1}^{j-2}\bar{k}_n(\bar{E}-\bar{\mathrm{v}}_n)d\right].
\end{aligned}
\tag{S29}
$$

Since $\bar{\mathrm{v}}_i$, $\bar{\mathrm{v}}_{i-1}$, $\bar{\mathrm{v}}_{j-1}$ and $\bar{\mathrm{v}}_j$ are independent random variables for $j > i+1$, we have

$$
\begin{aligned}
&\left\langle(\bar{\mathrm{v}}_i-\bar{\mathrm{v}}_{i-1})(\bar{\mathrm{v}}_j-\bar{\mathrm{v}}_{j-1})e^{2i\Phi_{i,j}}\right\rangle_c \\
&\approx\left\langle(\bar{\mathrm{v}}_i-\bar{\mathrm{v}}_{i-1})(\bar{\mathrm{v}}_j-\bar{\mathrm{v}}_{j-1})\left[1-2i\left(k_{x0}+\frac{k_y^2}{k_{x0}}\right)d\,\frac{\bar{\mathrm{v}}_i+\bar{\mathrm{v}}_{j-1}}{\bar{E}}\right]\right\rangle_c\left\langle\exp\left[2i\sum_{n=i+1}^{j-2}\bar{k}_n(\bar{E}-\bar{\mathrm{v}}_n)d\right]\right\rangle_c e^{4ik_{x0}d} \\
&=\left[\left\langle(\bar{\mathrm{v}}_i-\bar{\mathrm{v}}_{i-1})(\bar{\mathrm{v}}_j-\bar{\mathrm{v}}_{j-1})\right\rangle_c-2i\left(k_{x0}+\frac{k_y^2}{k_{x0}}\right)\frac{d}{\bar{E}}\left\langle(\bar{\mathrm{v}}_i-\bar{\mathrm{v}}_{i-1})(\bar{\mathrm{v}}_j-\bar{\mathrm{v}}_{j-1})(\bar{\mathrm{v}}_i+\bar{\mathrm{v}}_{j-1})\right\rangle_c\right] \\
&\cdot\left\langle\exp\left[2i\sum_{n=i+1}^{j-2}\bar{k}_n(\bar{E}-\bar{\mathrm{v}}_n)d\right]\right\rangle_c e^{4ik_{x0}d} = 0.
\end{aligned}
\tag{S30}
$$

Here

$$
\left\langle(\bar{\mathrm{v}}_i-\bar{\mathrm{v}}_{i-1})(\bar{\mathrm{v}}_j-\bar{\mathrm{v}}_{j-1})\right\rangle_c = \left(\langle\bar{\mathrm{v}}_i\rangle_c-\langle\bar{\mathrm{v}}_{i-1}\rangle_c\right)\left(\langle\bar{\mathrm{v}}_j\rangle_c-\langle\bar{\mathrm{v}}_{j-1}\rangle_c\right) = 0,
$$

and

$$
\begin{aligned}
&\left\langle(\bar{\mathrm{v}}_i-\bar{\mathrm{v}}_{i-1})(\bar{\mathrm{v}}_j-\bar{\mathrm{v}}_{j-1})(\bar{\mathrm{v}}_i+\bar{\mathrm{v}}_{j-1})\right\rangle_c \\
&=\left\langle\bar{\mathrm{v}}_i^2-\bar{\mathrm{v}}_{i-1}\bar{\mathrm{v}}_i\right\rangle_c\left(\langle\bar{\mathrm{v}}_j\rangle_c-\langle\bar{\mathrm{v}}_{j-1}\rangle_c\right)+\left\langle\bar{\mathrm{v}}_{j-1}\bar{\mathrm{v}}_j-\bar{\mathrm{v}}_{j-1}^2\right\rangle_c\left(\langle\bar{\mathrm{v}}_i\rangle_c-\langle\bar{\mathrm{v}}_{i-1}\rangle_c\right) \\
&=0.
\end{aligned}
$$

For $j = i+1$, the phase factor is written as

$$e^{2i\Phi_{i,j}} = e^{2i\Phi_{i,i+1}} = e^{2i\bar{k}_i(\bar{E}-\bar{v}_i)d} \approx e^{2ik_{x0}d}\left[1-2i\left(k_{x0}+\frac{k_y^2}{k_{x0}}\right)d\,\frac{\bar{v}_i}{\bar{E}}\right]. \tag{S31}$$

And the ensemble average of $\left(\bar{v}_i - \bar{v}_{i-1}\right)\left(\bar{v}_j - \bar{v}_{j-1}\right)e^{2i\Phi_{i,j}}$ is expressed as follows,

$$\begin{aligned}
&\left\langle \left(\bar{v}_i - \bar{v}_{i-1}\right)\left(\bar{v}_j - \bar{v}_{j-1}\right)e^{2i\Phi_{i,j}} \right\rangle_c = \left\langle \left(\bar{v}_i - \bar{v}_{i-1}\right)\left(\bar{v}_{i+1} - \bar{v}_i\right)e^{2i\Phi_{i,i+1}} \right\rangle_c \\
&\approx e^{2ik_{x0}d}\left\langle \bar{v}_i\bar{v}_{i+1} - \bar{v}_i^2 - \bar{v}_{i-1}\bar{v}_{i+1} + \bar{v}_{i-1}\bar{v}_i \right\rangle_c \\
&-2ie^{2ik_{x0}d}\left(k_{x0}+\frac{k_y^2}{k_{x0}}\right)\frac{d}{\bar{E}}\left\langle \bar{v}_i^2\bar{v}_{i+1} - \bar{v}_i^3 - \bar{v}_{i-1}\bar{v}_i\bar{v}_{i+1} + \bar{v}_{i-1}\bar{v}_i^2 \right\rangle_c \\
&= -\frac{1}{3}\bar{W}^2 e^{2ik_{x0}d}.
\end{aligned} \tag{S32}$$

Here all ensemble averages are zero except $\left\langle \bar{v}_i^2 \right\rangle_c = \frac{1}{3}\bar{W}^2$. So we can obtain $\kappa(\bar{E}, k_y)$ from Eq. (S28) as

$$\kappa(\bar{E}, k_y) \approx -\frac{4\bar{W}^2}{3\bar{E}^6} e^{2ik_{x0}d}. \tag{S33}$$

Thus, we can get $\alpha_1 = 2\Re(\kappa) = -\frac{8\bar{W}^2}{3\bar{E}^6}\cos(2k_{x0}d)$, and the Lyapunov exponent from Eq. (S26) for $\bar{W} << \bar{E}$ and small $k_y$ as,

$$\gamma \approx \frac{\bar{W}^2 k_y^4}{12\bar{E}^6 d}\left[1-\cos\left(2k_{x0}d\right)\right] = \frac{\bar{W}^2 \sin^4\theta}{12\bar{E}^2 d}\left[1-\cos\left(2\bar{E}d\cos\theta\right)\right]. \tag{S34}$$

Here $k_{x0} = \sqrt{\bar{E}^2 - k_y^2} = \bar{E}\cos\theta$.

For pseudospin-1/2 systems, the reflection amplitudes for $\bar{W} << \bar{E}$ and small $k_y$ can be approximated as

$$r_{n-1,n} \approx -\frac{U_n}{2(\bar{E}-\bar{\mathrm{v}}_{n-1})}k_y^2 + \frac{1}{2}iU_n k_y,$$
$$r_{n-1,n} \approx \frac{U_n}{2(\bar{E}-\bar{\mathrm{v}}_n)}k_y^2 + \frac{1}{2}iU_n k_y. \tag{S35}$$

Thus, $\gamma_2$ for pseudospin-1/2 systems can be written as

$$\begin{aligned}\gamma_2 &\approx -\frac{1}{2d}\left\langle \ln\left|1-r_{n,n-1}\right| + \ln\left|1-r_{n-1,n}\right|\right\rangle_c = -\frac{1}{2d}\left\langle \ln\left|1-(r_{n,n-1}+r_{n-1,n})+r_{n,n-1}r_{n-1,n}\right|\right\rangle_c \\ &\approx -\frac{1}{2d}\left\langle \ln\left(1-\frac{1}{4}U_n^2k_y^2\right)\right\rangle_c \approx \frac{1}{8d}k_y^2\left\langle U_n^2\right\rangle_c \\ &\approx \frac{1}{8d}k_y^2\frac{1}{\bar{E}^4}\left\langle\left(\bar{\mathrm{v}}_n-\bar{\mathrm{v}}_{n-1}\right)^2\right\rangle_c = \frac{1}{12\bar{E}^2 d}\bar{W}^2\sin^2\theta.\end{aligned} \tag{S36}$$

For $\gamma_1$ in pseudospin-1/2 systems, it also can be obtained using the Leibniz formula in a way similar to pseudospin-1 systems,

$$\gamma_1 \approx -\frac{\bar{W}^2\sin^2\theta}{12\bar{E}^2}\cos\left(2\bar{E}d\cos\theta\right). \tag{S37}$$

Thus, the Lyapunov exponent in pseudospin-1/2 systems can be expressed as follows,

$$\gamma = \xi^{-1} = \gamma_1 + \gamma_2 \approx \frac{\bar{W}^2\sin^2\theta}{12\bar{E}^2 d}\left[1-\cos\left(2\bar{E}d\cos\theta\right)\right]. \tag{S38}$$

From Eqs. (S34) and (S38), we can see that the two pseudospin systems have different $\theta$-dependences of the Lyapunov exponent.

### $\gamma_2$ for Pseudospin-1 Systems in the Case $\bar{W} > \bar{E}$

We first consider the term $\left\langle \ln|e^{2i\Phi_{1,N+1}}|\right\rangle_c$ with $\Phi_{1,N+1} = \sum_{n=1}^{N}\bar{k}_n(\bar{E}-\bar{\mathrm{v}}_n)d$ . In the case of $\bar{W} > \bar{E}$,

$\bar{k}_n = \sqrt{1-\frac{k_y^2}{(\bar{E}-\bar{\mathrm{v}}_n)^2}}$ is imaginary when $\left|\bar{E}-\bar{\mathrm{v}}_n\right| < |k_y|$ and real when $\left|\bar{E}-\bar{\mathrm{v}}_n\right| \geq |k_y|$ . Thus, we have

$$\left\langle \ln|e^{2i\Phi_{1,N+1}}|\right\rangle_c = \left\langle \ln\left|\exp\left[2i\sum_{n=1}^{N}\bar{k}_n(\bar{E}-\bar{\mathrm{v}}_n)d\right]\right|\right\rangle_c$$
$$= -2d\sum_n \left\langle (\bar{E}-\bar{\mathrm{v}}_n)\sqrt{\frac{k_y^2}{(\bar{E}-\bar{\mathrm{v}}_n)^2}-1}\right\rangle_c, \qquad \left(\left|\bar{E}-\bar{\mathrm{v}}_n\right| < |k_y|\right). \tag{S39}$$

Since the distribution of $\bar{\mathrm{v}}_n$ is uniform, we find

$$\left\langle (\bar{E}-\bar{\mathrm{v}}_n)\sqrt{\frac{k_y^2}{(\bar{E}-\bar{\mathrm{v}}_n)^2}-1}\right\rangle_c \propto \int_{\bar{E}-|k_y|}^{\bar{E}+|k_y|} d\bar{\mathrm{v}}_n (\bar{E}-\bar{\mathrm{v}}_n)\sqrt{\frac{k_y^2}{(\bar{E}-\bar{\mathrm{v}}_n)^2}-1}$$
$$= -\int_{-|k_y|}^{|k_y|} x\sqrt{\frac{k_y^2}{x^2}-1}dx = 0, \tag{S40}$$

for sufficiently small $k_y$ satisfying $|k_y| < \bar{W}-\bar{E}$. Thus, $\left\langle \ln|e^{2i\Phi_{1,N+1}}|\right\rangle_c$ does not have contributions to $\gamma_2$ in the limit of $|k_y| < \bar{W}-\bar{E}$ and $\gamma_2$ can be written as

$$\gamma_2 \approx -\frac{1}{2d}\left\langle \ln\left|1-r_{n,n-1}^2\right|\right\rangle_c = -\frac{1}{8\bar{W}^2 d}\int_{-\bar{W}}^{\bar{W}} d\bar{\mathrm{v}}_n \int_{-\bar{W}}^{\bar{W}} d\bar{\mathrm{v}}_{n-1} \ln\left|\frac{4\sqrt{1-\frac{k_y^2}{(\bar{E}-\bar{\mathrm{v}}_n)^2}}\sqrt{1-\frac{k_y^2}{(\bar{E}-\bar{\mathrm{v}}_{n-1})^2}}}{\left(\sqrt{1-\frac{k_y^2}{(\bar{E}-\bar{\mathrm{v}}_n)^2}}+\sqrt{1-\frac{k_y^2}{(\bar{E}-\bar{\mathrm{v}}_{n-1})^2}}\right)^2}\right|$$
$$= -\frac{1}{8\bar{W}^2 d}\int_{-\bar{W}}^{\bar{W}} d\bar{\mathrm{v}}_n \int_{-\bar{W}}^{\bar{W}} d\bar{\mathrm{v}}_{n-1}\left[\ln 4 + \frac{1}{2}\ln\left|1-\frac{k_y^2}{(\bar{E}-\bar{\mathrm{v}}_n)^2}\right| + \frac{1}{2}\ln\left|1-\frac{k_y^2}{(\bar{E}-\bar{\mathrm{v}}_{n-1})^2}\right|\right. \tag{S41}$$
$$\left. -2\ln\left|\sqrt{1-\frac{k_y^2}{(\bar{E}-\bar{\mathrm{v}}_n)^2}}+\sqrt{1-\frac{k_y^2}{(\bar{E}-\bar{\mathrm{v}}_{n-1})^2}}\right|\right].$$

For convenience, we separate the integral space into the following 9 parts:

(1) $|\bar{E}-\bar{\mathrm{v}}_n| \le |k_y|$ and $|\bar{E}-\bar{\mathrm{v}}_{n-1}| \le |k_y|$

$$I_1 = -\frac{1}{8\bar{W}^2 d}\int_{\bar{E}-|k_y|}^{\bar{E}+|k_y|} d\bar{v}_n \int_{\bar{E}-|k_y|}^{\bar{E}+|k_y|} d\bar{v}_{n-1}\left[\ln 4 + \frac{1}{2}\ln\left(\frac{k_y^2}{(\bar{E}-\bar{v}_n)^2}-1\right)+\frac{1}{2}\ln\left(\frac{k_y^2}{(\bar{E}-\bar{v}_{n-1})^2}-1\right)\right.$$
$$\left.-2\ln\left(\sqrt{\frac{k_y^2}{(\bar{E}-\bar{v}_n)^2}-1}+\sqrt{\frac{k_y^2}{(\bar{E}-\bar{v}_{n-1})^2}-1}\right)\right]$$
$$= -\frac{2k_y^2 \ln 2}{\bar{W}^2 d} + \frac{1}{4\bar{W}^2 d}\int_{\bar{E}-|k_y|}^{\bar{E}+|k_y|} d\bar{v}_n \int_{\bar{E}-|k_y|}^{\bar{E}+|k_y|} d\bar{v}_{n-1} \ln\left(\sqrt{\frac{k_y^2}{(\bar{E}-\bar{v}_n)^2}-1}+\sqrt{\frac{k_y^2}{(\bar{E}-\bar{v}_{n-1})^2}-1}\right). \quad \text{(S42)}$$

To calculate $I_1$, we use the following variable substitutions,

$$x_n = \frac{|k_y|}{\bar{E}-\bar{v}_n}, \qquad x_{n-1} = \frac{|k_y|}{\bar{E}-\bar{v}_{n-1}},$$
$$d\bar{v}_n = \frac{|k_y|}{x_n^2} dx_n, \qquad d\bar{v}_{n-1} = \frac{|k_y|}{x_{n-1}^2} dx_{n-1}. \quad \text{(S43)}$$

Then $I_1$ can be rewritten as follows,

$$I_1 = -\frac{2k_y^2 \ln 2}{\bar{W}^2 d} + \frac{k_y^2}{\bar{W}^2 d}\int_1^\infty \frac{dx_n}{x_n^2}\int_1^\infty \frac{dx_{n-1}}{x_{n-1}^2} \ln\left(\sqrt{x_n^2-1}+\sqrt{x_{n-1}^2-1}\right)$$
$$= -\frac{2k_y^2 \ln 2}{\bar{W}^2 d} + \frac{k_y^2}{\bar{W}^2 d}(1+\ln 2) = \frac{k_y^2}{\bar{W}^2 d}(1-\ln 2). \quad \text{(S44)}$$

(2) $\bar{v}_n - \bar{E} > |k_y|$ and $\bar{v}_{n-1} - \bar{E} > |k_y|$

$$I_2 = -\frac{1}{8\bar{W}^2 d}\int_{\bar{E}+|k_y|}^{\bar{W}} d\bar{v}_n \int_{\bar{E}+|k_y|}^{\bar{W}} d\bar{v}_{n-1}\left[\ln 4 + \frac{1}{2}\ln\left(1-\frac{k_y^2}{(\bar{E}-\bar{v}_n)^2}\right)+\frac{1}{2}\ln\left(1-\frac{k_y^2}{(\bar{E}-\bar{v}_{n-1})^2}\right)\right.$$
$$\left.-2\ln\left(\sqrt{1-\frac{k_y^2}{(\bar{E}-\bar{v}_n)^2}}+\sqrt{1-\frac{k_y^2}{(\bar{E}-\bar{v}_{n-1})^2}}\right)\right]$$

$$I_2 = -\frac{(\bar{W}-\bar{E}-|k_y|)^2 \ln 2}{4\bar{W}^2 d} - \frac{\bar{W}-\bar{E}-|k_y|}{8\bar{W}^2 d}\left[(\bar{W}-\bar{E})\ln\left(1-\frac{k_y^2}{(\bar{W}-\bar{E})^2}\right)\right.$$

$$\left.+|k_y|\ln\frac{\bar{W}-\bar{E}+|k_y|}{\bar{W}-\bar{E}-|k_y|} - 2|k_y|\ln 2\right] +$$

$$\frac{k_y^2}{4\bar{W}^2 d}\int_{\frac{|k_y|}{\bar{W}-\bar{E}}}^{1}\frac{dx_n}{x_n^2}\int_{\frac{|k_y|}{\bar{W}-\bar{E}}}^{1}\frac{dx_{n-1}}{x_{n-1}^2}\ln\left(\sqrt{1-x_n^2}+\sqrt{1-x_{n-1}^2}\right)$$

$$= -\frac{(\bar{W}-\bar{E}-|k_y|)^2 \ln 2}{4\bar{W}^2 d} - \frac{\bar{W}-\bar{E}-|k_y|}{8\bar{W}^2 d}\left[(\bar{W}-\bar{E})\ln\left(1-\frac{k_y^2}{(\bar{W}-\bar{E})^2}\right)\right.$$

$$\left.+|k_y|\ln\frac{\bar{W}-\bar{E}+|k_y|}{\bar{W}-\bar{E}-|k_y|} - 2|k_y|\ln 2\right] +$$

$$\frac{(\bar{W}-\bar{E})^2}{16\bar{W}^2 d}\left\{-2\ln\frac{\bar{W}-\bar{E}+|k_y|}{\bar{W}-\bar{E}-|k_y|} + \frac{4|k_y|}{\bar{W}-\bar{E}}\left[\frac{(1+\ln 2)|k_y|}{\bar{W}-\bar{E}} - 1\right] + 4\ln 2\right.$$

$$\left.+\left(\frac{|k_y|}{\bar{W}-\bar{E}}-2\right)\frac{2|k_y|}{\bar{W}-\bar{E}}\ln\left(1-\frac{|k_y|}{\bar{W}-\bar{E}}\right) - \frac{2|k_y|}{\bar{W}-\bar{E}}\left(2+\frac{|k_y|}{\bar{W}-\bar{E}}\right)\ln\left(1+\frac{|k_y|}{\bar{W}-\bar{E}}\right)\right\}. \qquad \text{(S45)}$$

For small $k_y$, i.e., $|k_y| << \bar{W}-\bar{E}$, the first two leading terms of $I_2$ are

$$I_2 \approx \frac{(-2+3\ln 2)(\bar{W}-\bar{E})|k_y|}{4\bar{W}^2 d} + (1-2\ln 2)\frac{k_y^2}{8\bar{W}^2 d}. \qquad \text{(S46)}$$

(3) $\bar{E}-\text{v}_n > |k_y|$ and $\bar{E}-\text{v}_{n-1} > |k_y|$

$$I_3 = -\frac{1}{8\bar{W}^2 d}\int_{-\bar{W}}^{\bar{E}-|k_y|} d\text{v}_n \int_{-\bar{W}}^{\bar{E}-|k_y|} d\text{v}_{n-1}\left[\ln 4 + \frac{1}{2}\ln\left(1-\frac{k_y^2}{(\bar{E}-\text{v}_n)^2}\right) + \frac{1}{2}\ln\left(1-\frac{k_y^2}{(\bar{E}-\text{v}_{n-1})^2}\right)\right.$$

$$\left. -2\ln\left(\sqrt{1-\frac{k_y^2}{(\bar{E}-\text{v}_n)^2}} + \sqrt{1-\frac{k_y^2}{(\bar{E}-\text{v}_{n-1})^2}}\right)\right]$$

$$= -\frac{(\bar{W}+\bar{E}-|k_y|)^2 \ln 2}{4\bar{W}^2 d} - \frac{\bar{W}+\bar{E}-|k_y|}{8\bar{W}^2 d}(\bar{W}+\bar{E})\ln\left(1-\frac{k_y^2}{(\bar{W}+\bar{E})^2}\right) \qquad \text{(S47)}$$

$$+|k_y|\ln\frac{\bar{W}+\bar{E}+|k_y|}{\bar{W}+\bar{E}-|k_y|} - 2|k_y|\ln 2\Bigg] +$$

$$\frac{k_y^2}{4\bar{W}^2 d}\int_{\frac{|k_y|}{\bar{W}+\bar{E}}}^{1}\frac{dx_n}{x_n^2}\int_{\frac{|k_y|}{\bar{W}+\bar{E}}}^{1}\frac{dx_{n-1}}{x_{n-1}^2}\ln\left(\sqrt{1-x_n^2}+\sqrt{1-x_{n-1}^2}\right)$$

Thus,

$$
\begin{aligned}
I_3 = &-\frac{(\bar{W}+\bar{E}-|k_y|)^2 \ln 2}{4\bar{W}^2 d} - \frac{\bar{W}+\bar{E}-|k_y|}{8\bar{W}^2 d}\Bigg[(\bar{W}+\bar{E})\ln\left(1-\frac{k_y^2}{(\bar{W}+\bar{E})^2}\right) \\
&+|k_y|\ln\frac{\bar{W}+\bar{E}+|k_y|}{\bar{W}+\bar{E}-|k_y|} - 2|k_y|\ln 2\Bigg] + \\
&\frac{(\bar{W}+\bar{E})^2}{16\bar{W}^2 d}\Bigg\{-2\ln\frac{\bar{W}+\bar{E}+|k_y|}{\bar{W}+\bar{E}-|k_y|} + \frac{4|k_y|}{\bar{W}+\bar{E}}\left[\frac{(1+\ln 2)|k_y|}{\bar{W}+\bar{E}} - 1\right] + 4\ln 2 \\
&+\left(\frac{|k_y|}{\bar{W}+\bar{E}} - 2\right)\frac{2|k_y|}{\bar{W}+\bar{E}}\ln\left(1-\frac{|k_y|}{\bar{W}+\bar{E}}\right) - \frac{2|k_y|}{\bar{W}+\bar{E}}\left(2+\frac{|k_y|}{\bar{W}+\bar{E}}\right)\ln\left(1+\frac{|k_y|}{\bar{W}+\bar{E}}\right)\Bigg\}.
\end{aligned}
\tag{S48}
$$

At small $k_y$, $I_3$ can be expressed as

$$
I_3 \approx \frac{(-2+3\ln 2)(\bar{W}+\bar{E})|k_y|}{4\bar{W}^2 d} + (1-2\ln 2)\frac{k_y^2}{8\bar{W}^2 d}. \tag{S49}
$$

(4) $\bar{E}-\bar{v}_n > |k_y|$ and $\bar{v}_{n-1}-\bar{E} > |k_y|$, or $\bar{E}-\bar{v}_{n-1} > |k_y|$ and $\bar{v}_n-\bar{E} > |k_y|$

$$
\begin{aligned}
I_4 = &-\frac{1}{8\bar{W}^2 d}\int_{-\bar{W}}^{\bar{E}-|k_y|} d\bar{v}_n \int_{\bar{E}+|k_y|}^{\bar{W}} d\bar{v}_{n-1}\Bigg[\ln 4 + \frac{1}{2}\ln\left(1-\frac{k_y^2}{(\bar{E}-\bar{v}_n)^2}\right) + \frac{1}{2}\ln\left(1-\frac{k_y^2}{(\bar{E}-\bar{v}_{n-1})^2}\right) \\
&-2\ln\left(\sqrt{1-\frac{k_y^2}{(\bar{E}-\bar{v}_n)^2}} + \sqrt{1-\frac{k_y^2}{(\bar{E}-\bar{v}_{n-1})^2}}\right)\Bigg],
\end{aligned}
\tag{S50}
$$

$$
\begin{aligned}
I_5 = &-\frac{1}{8\bar{W}^2 d}\int_{\bar{E}+|k_y}^{\bar{W}} d\bar{v}_n \int_{-\bar{W}}^{\bar{E}-|k_y|} d\bar{v}_{n-1}\Bigg[\ln 4 + \frac{1}{2}\ln\left(1-\frac{k_y^2}{(\bar{E}-\bar{v}_n)^2}\right) + \frac{1}{2}\ln\left(1-\frac{k_y^2}{(\bar{E}-\bar{v}_{n-1})^2}\right) \\
&-2\ln\left(\sqrt{1-\frac{k_y^2}{(\bar{E}-\bar{v}_n)^2}} + \sqrt{1-\frac{k_y^2}{(\bar{E}-\bar{v}_{n-1})^2}}\right)\Bigg].
\end{aligned}
\tag{S51}
$$

Note that the integrands for $I_4$ and $I_5$ are symmetric after interchanging the variables $\bar{v}_n$ and $\bar{v}_{n-1}$, we have $I_4 = I_5$.

$$I_4 = -\frac{(\bar{W}-|k_y|)^2-\bar{E}^2}{4\bar{W}^2 d}\ln 2 - \frac{\bar{W}-\bar{E}-|k_y|}{16\bar{W}^2 d}\left[(\bar{W}+\bar{E})\ln\left(1-\frac{k_y^2}{(\bar{W}+\bar{E})^2}\right)-2|k_y|\ln 2\right.$$
$$\left.+|k_y|\ln\frac{\bar{W}+\bar{E}+|k_y|}{\bar{W}+\bar{E}-|k_y|}\right] - \frac{\bar{W}+\bar{E}-|k_y|}{16\bar{W}^2 d}\left[(\bar{W}-\bar{E})\ln\left(1-\frac{k_y^2}{(\bar{W}-\bar{E})^2}\right)-2|k_y|\ln 2\right. \tag{S52}$$
$$\left.+|k_y|\ln\frac{\bar{W}-\bar{E}+|k_y|}{\bar{W}-\bar{E}-|k_y|}\right] + \frac{k_y^2}{4\bar{W}^2 d}\int_{\frac{|k_y|}{\bar{W}+\bar{E}}}^{1}\frac{dx_n}{x_n^2}\int_{\frac{|k_y|}{\bar{W}-\bar{E}}}^{1}\frac{dx_{n-1}}{x_{n-1}^2}\ln\left(\sqrt{1-x_n^2}+\sqrt{1-x_{n-1}^2}\right).$$

For the last integral, we have

$$I = \int_a^1 \frac{dx_n}{x_n^2}\int_b^1 \frac{dx_{n-1}}{x_{n-1}^2}\ln\left(\sqrt{1-x_n^2}+\sqrt{1-x_{n-1}^2}\right)$$
$$= \frac{1}{4a^2b^2}\left\{2ab-2a^2b-2ab^2+2a^2b^2-2ab\sqrt{(1-a^2)(1-b^2)}-(a^2-2ab^2+a^2b^2)\ln\frac{1+b}{1-b}\right.$$
$$-4ab^2\ln 2+4a^2b^2\ln 2+b^2\ln(1-a)+a^2b^2\ln\frac{1-a}{1+a}-b^2\ln(1+a)+2ab^2\ln\frac{4}{1-a^2}$$
$$+2ab^2\ln(1-b)-2ab^2\ln(1+b)+2ab\ln(1-b^2)-2a^2b\ln(1-b^2) \tag{S53}$$
$$-a^2\ln\frac{(a-b)(b\sqrt{1-a^2}+a\sqrt{1-b^2})}{(a+b)(-b\sqrt{1-a^2}+a\sqrt{1-b^2})}+4ab\ln\left(1+\sqrt{\frac{1-a^2}{1-b^2}}\right)$$
$$\left.+2b^2\ln\frac{b\sqrt{1-a^2}-a\sqrt{1-b^2}}{b\sqrt{1-a^2}+a\sqrt{1-b^2}}-b^2\ln\frac{1-ab-\sqrt{(1-a^2)(1-b^2)}}{1+ab-\sqrt{(1-a^2)(1-b^2)}}\right\},$$

where $a=|k_y|/(\bar{W}+\bar{E})$, $b=|k_y|/(\bar{W}-\bar{E})$ and $0<a<b<1$. At small $k_y$, $I_4$ and $I_5$ can be expressed as

$$I_4 = I_5 \approx (-2+3\ln 2)\frac{|k_y|}{4\bar{W}d}+(1-2\ln 2)\frac{k_y^2}{8\bar{W}^2 d}. \tag{S54}$$

(5) $\left|\bar{E}-\bar{v}_n\right|\le|k_y|$ and $\bar{v}_{n-1}-\bar{E}>|k_y|$, or $\left|\bar{E}-\bar{v}_{n-1}\right|\le|k_y|$ and $\bar{v}_n-\bar{E}>|k_y|$

$$I_6 = -\frac{1}{8\bar{W}^2 d}\int_{\bar{E}-|k_y|}^{\bar{E}+|k_y|}d\bar{v}_n\int_{\bar{E}+|k_y|}^{\bar{W}}d\bar{v}_{n-1}\left[\ln 4+\frac{1}{2}\ln\left(\frac{k_y^2}{(\bar{E}-\bar{v}_n)^2}-1\right)+\frac{1}{2}\ln\left(1-\frac{k_y^2}{(\bar{E}-\bar{v}_{n-1})^2}\right)\right.$$
$$\left.-\ln\left(\frac{k_y^2}{(\bar{E}-\bar{v}_n)^2}-\frac{k_y^2}{(\bar{E}-\bar{v}_{n-1})^2}\right)\right], \tag{S55}$$

$$I_7 = -\frac{1}{8\bar{W}^2 d}\int_{\bar{E}+|k_y|}^{\bar{W}} d\bar{v}_n \int_{\bar{E}-|k_y|}^{\bar{E}+|k_y|} d\bar{v}_{n-1}\left[\ln 4 + \frac{1}{2}\ln\left(1-\frac{k_y^2}{(\bar{E}-\bar{v}_n)^2}\right)+\frac{1}{2}\ln\left(\frac{k_y^2}{(\bar{E}-\bar{v}_{n-1})^2}-1\right)\right.$$
$$\left.-\ln\left(\frac{k_y^2}{(\bar{E}-\bar{v}_{n-1})^2}-\frac{k_y^2}{(\bar{E}-\bar{v}_n)^2}\right)\right]. \tag{S56}$$

It is straightforward to show $I_6 = I_7$. After integration, $I_7$ can be expressed as

$$I_7 = -\frac{(\bar{W}-\bar{E}-|k_y|)|k_y|}{2\bar{W}^2 d}\ln 2 - \frac{(\bar{W}-\bar{E})|k_y|-k_y^2}{4\bar{W}^2 d}\ln 2 - \frac{|k_y|}{8\bar{W}^2 d}\left[(\bar{W}-\bar{E})\ln\left(1-\frac{k_y^2}{(\bar{W}-\bar{E})^2}\right)\right.$$
$$\left.+|k_y|\ln\frac{\bar{W}-\bar{E}+|k_y|}{\bar{W}-\bar{E}-|k_y|}-2|k_y|\ln 2\right]+\frac{k_y^2}{4\bar{W}^2 d}\int_{\frac{|k_y|}{\bar{W}-\bar{E}}}^{1}\frac{dx_n}{x_n^2}\int_1^{\infty}\frac{dx_{n-1}}{x_{n-1}^2}\ln(x_{n-1}^2-x_n^2)$$
$$= -\frac{(\bar{W}-\bar{E}-|k_y|)|k_y|}{2\bar{W}^2 d}\ln 2 - \frac{(\bar{W}-\bar{E})|k_y|-k_y^2}{4\bar{W}^2 d}\ln 2 - \frac{|k_y|}{8\bar{W}^2 d}\left[(\bar{W}-\bar{E})\ln\left(1-\frac{k_y^2}{(\bar{W}-\bar{E})^2}\right)\right. \tag{S57}$$
$$\left.+|k_y|\ln\frac{\bar{W}-\bar{E}+|k_y|}{\bar{W}-\bar{E}-|k_y|}-2|k_y|\ln 2\right]-\frac{k_y^2}{4\bar{W}^2 d}+\frac{\bar{W}-\bar{E}}{4\bar{W}^2 d}|k_y|+\frac{k_y^2}{8\bar{W}^2 d}\ln\frac{\bar{W}-\bar{E}+|k_y|}{\bar{W}-\bar{E}-|k_y|}$$
$$+\frac{(\bar{W}-\bar{E})^2}{8\bar{W}^2 d}\ln\frac{\bar{W}-\bar{E}+|k_y|}{\bar{W}-\bar{E}-|k_y|}-\frac{k_y^2\ln 2}{2\bar{W}^2 d}+\frac{\bar{W}-\bar{E}}{4\bar{W}^2 d}|k_y|\ln\left(1-\frac{k_y^2}{(\bar{W}-\bar{E})^2}\right).$$

At small $k_y$, Eq. (S57) gives

$$I_7 \approx (2-3\ln 2)\frac{(\bar{W}-\bar{E})|k_y|}{4\bar{W}^2 d}+\frac{k_y^2}{4\bar{W}^2 d}(2\ln 2-1). \tag{S58}$$

(6) $\left|\bar{E}-\bar{v}_n\right| \le |k_y|$ and $\bar{E}-\bar{v}_{n-1} > |k_y|$, or $\left|\bar{E}-\bar{v}_{n-1}\right| \le |k_y|$ and $\bar{E}-\bar{v}_n > |k_y|$

$$I_8 = -\frac{1}{8\bar{W}^2 d}\int_{-\bar{W}}^{\bar{E}-|k_y|} d\bar{v}_n \int_{\bar{E}-|k_y|}^{\bar{E}+|k_y|} d\bar{v}_{n-1}\left[\ln 4 + \frac{1}{2}\ln\left(1-\frac{k_y^2}{(\bar{E}-\bar{v}_n)^2}\right)+\frac{1}{2}\ln\left(\frac{k_y^2}{(\bar{E}-\bar{v}_{n-1})^2}-1\right)\right.$$
$$\left.-\ln\left(\frac{k_y^2}{(\bar{E}-\bar{v}_{n-1})^2}-\frac{k_y^2}{(\bar{E}-\bar{v}_n)^2}\right)\right], \tag{S59}$$

$$I_9 = -\frac{1}{8\bar{W}^2 d}\int_{\bar{E}-|k_y|}^{\bar{E}+|k_y|} d\bar{v}_n \int_{-\bar{W}}^{\bar{E}-|k_y|} d\bar{v}_{n-1}\left[\ln 4 + \frac{1}{2}\ln\left(\frac{k_y^2}{(\bar{E}-\bar{v}_n)^2}-1\right)+\frac{1}{2}\ln\left(1-\frac{k_y^2}{(\bar{E}-\bar{v}_{n-1})^2}\right)\right.$$
$$\left.-\ln\left(\frac{k_y^2}{(\bar{E}-\bar{v}_n)^2}-\frac{k_y^2}{(\bar{E}-\bar{v}_{n-1})^2}\right)\right]. \tag{S60}$$

Also, we have $I_8 = I_9$. $I_8$ can be integrated as

$$I_8 = -\frac{(\bar{W}+\bar{E}-|k_y|)|k_y|}{2\bar{W}^2 d}\ln 2 - \frac{(\bar{W}+\bar{E})|k_y|-k_y^2}{4\bar{W}^2 d}\ln 2 - \frac{|k_y|}{8\bar{W}^2 d}\left[(\bar{W}+\bar{E})\ln\left(1-\frac{k_y^2}{(\bar{W}+\bar{E})^2}\right)\right.$$
$$\left. +|k_y|\ln\frac{\bar{W}+\bar{E}+|k_y|}{\bar{W}+\bar{E}-|k_y|} - 2|k_y|\ln 2\right] + \frac{k_y^2}{4\bar{W}^2 d}\int_{\frac{|k_y|}{\bar{W}+\bar{E}}}^{1}\frac{dx_n}{x_n^2}\int_1^{\infty}\frac{dx_{n-1}}{x_{n-1}^2}\ln(x_{n-1}^2 - x_n^2)$$
$$= -\frac{(\bar{W}+\bar{E}-|k_y|)|k_y|}{2\bar{W}^2 d}\ln 2 - \frac{(\bar{W}+\bar{E})|k_y|-k_y^2}{4\bar{W}^2 d}\ln 2 - \frac{|k_y|}{8\bar{W}^2 d}\left[(\bar{W}+\bar{E})\ln\left(1-\frac{k_y^2}{(\bar{W}+\bar{E})^2}\right)\right. \tag{S61}$$
$$\left. +|k_y|\ln\frac{\bar{W}+\bar{E}+|k_y|}{\bar{W}+\bar{E}-|k_y|} - 2|k_y|\ln 2\right] - \frac{k_y^2}{4\bar{W}^2 d} + \frac{\bar{W}+\bar{E}}{4\bar{W}^2 d}|k_y| + \frac{k_y^2}{8\bar{W}^2 d}\ln\frac{\bar{W}+\bar{E}+|k_y|}{\bar{W}+\bar{E}-|k_y|}$$
$$+\frac{(\bar{W}+\bar{E})^2}{8\bar{W}^2 d}\ln\frac{\bar{W}+\bar{E}+|k_y|}{\bar{W}+\bar{E}-|k_y|} - \frac{k_y^2\ln 2}{2\bar{W}^2 d} + \frac{\bar{W}+\bar{E}}{4\bar{W}^2 d}|k_y|\ln\left(1-\frac{k_y^2}{(\bar{W}+\bar{E})^2}\right).$$

For small $k_y$, Eq. (S61) gives

$$I_8 \approx (2-3\ln 2)\frac{(\bar{W}+\bar{E})|k_y|}{4\bar{W}^2 d} + \frac{k_y^2}{4\bar{W}^2 d}(2\ln 2 - 1). \tag{S62}$$

Summing over all these 9 integrals, we obtain $\gamma_2$ in pseudospin-1 systems,

$$\begin{aligned}
\gamma_2 = \sum_{i=1}^{9} I_i &\approx \frac{k_y^2}{\bar{W}^2 d}(1-\ln 2) + \frac{(-2+3\ln 2)(\bar{W}-\bar{E})|k_y|}{4\bar{W}^2 d} + (1-2\ln 2)\frac{k_y^2}{8\bar{W}^2 d} \\
&+ \frac{(-2+3\ln 2)(\bar{W}+\bar{E})|k_y|}{4\bar{W}^2 d} + (1-2\ln 2)\frac{k_y^2}{8\bar{W}^2 d} \\
&+2\times\left[(-2+3\ln 2)\frac{|k_y|}{4\bar{W}d} + (1-2\ln 2)\frac{k_y^2}{8\bar{W}^2 d}\right] \\
&+2\times\left[(2-3\ln 2)\frac{(\bar{W}-\bar{E})|k_y|}{4\bar{W}^2 d} + \frac{k_y^2}{4\bar{W}^2 d}(2\ln 2 - 1)\right] \\
&+2\times\left[(2-3\ln 2)\frac{(\bar{W}+\bar{E})|k_y|}{4\bar{W}^2 d} + \frac{k_y^2}{4\bar{W}^2 d}(2\ln 2 - 1)\right] \\
&= \frac{k_y^2}{2\bar{W}^2 d} = \frac{\bar{E}^2\sin^2\theta}{2\bar{W}^2 d}.
\end{aligned} \tag{S63}$$

## Figure Legends for Figures in Supporting Information

Fig. S1 Comparison of the localization length calculated by using the TMM method and analytical results shown in Eq. (18). Both $\bar{E}$ and $\bar{W}$ are in unit of $2\pi / d$ .

## Figures in Supporting Information

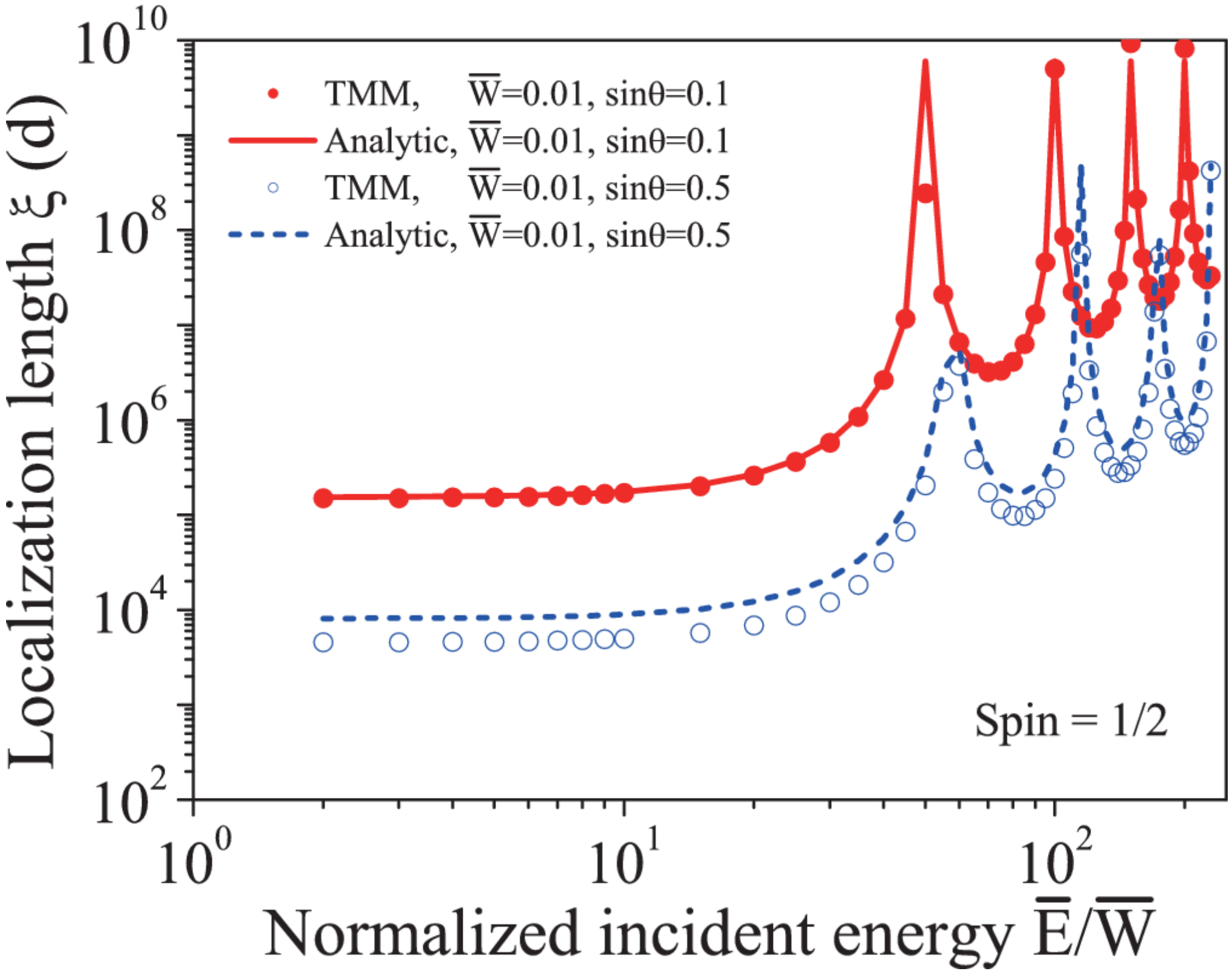

**Figure S1**